\documentclass{article}

\usepackage{amsmath}
\usepackage{openbib,epsfig,aces}
 
\begin{document}

\title{
Dynamical system analysis and forecasting of deformation produced by an
earthquake fault
}

\author{Marian Anghel\affiln{1}, Yehuda Ben-Zion\affiln{2}, 
and Ramiro Rico-Martinez\affiln{3}}

\affil{
  (1) Computer and Computational Sciences Division, Los Alamos National
  Laboratory, Los Alamos, NM, U. S. A (e-mail: manghel@lanl.gov).
  (2) Department of Earth Sciences, University of Southern California,
  Los Angeles, CA, U. S. A. (e-mail: benzion@usc.edu). 
  (3) Department of Chemical Engineering,
  Instituto Tecnologico de Celaya, 
  Celaya, Guanajuato, Mexico (e-mail: ramiro@losalamos.princeton.edu).
}

\maketitle

\begin{abstract}

  We present a method of constructing low-dimensional nonlinear models
  describing the main dynamical features of a discrete 2D cellular
  fault zone, with many degrees of freedom, embedded in a 3D elastic
  solid.  A given fault system is characterized by a set of parameters
  that describe the dynamics, rheology, property disorder, and fault
  geometry.  Depending on the location in the system parameter space
  we show that the coarse dynamics of the fault can be confined to an
  attractor whose dimension is significantly smaller than the space in
  which the dynamics takes place. Our strategy of system reduction is
  to search for a few coherent structures that dominate the dynamics
  and to capture the interaction between these coherent
  structures. The identification of the basic interacting structures
  is obtained by applying the Proper Orthogonal Decomposition (POD) to
  the surface deformations fields that accompany strike-slip faulting
  accumulated over equal time intervals.  We use a feed-forward
  artificial neural network (ANN) architecture for the identification
  of the system dynamics projected onto the subspace (model space)
  spanned by the most energetic coherent structures.  The ANN is
  trained using a standard back-propagation algorithm to predict (map)
  the values of the observed model state at a future time given the
  observed model state at the present time. This ANN provides an
  approximate, large scale, dynamical model for the fault. The map can
  be evaluated once to provide short term predictions or iterated to
  obtain prediction for the long term fault dynamics.

\end{abstract}

\section{1 Introduction}

A first principles approach to modeling and forecasting the dynamics
of an earthquake fault is not feasible at present because the
governing physical laws, geometric and structural fault properties,
and controlling variables (fault stresses and slips) are not fully
available. A practical alternative  is to build ``phenomenological''
models that attempt to estimate the overall character of the system's
dynamics.  These models quantify basic deterministic or stochastic
relationships involving only a few irreducible degrees of freedom,
which may be used for short-term prediction. We show that for earthquake
forecasting this can be done using the spatio-temporal strain
patterns embedded in the observable surface displacements.  Such an
approach is based on the observation that the large scale dynamics of
the system often evolves on a manifold (or invariant measure for the
case of strange attractors) with a dimension that is significantly
smaller than that of the system's phase space.  In such cases, a few
macroscopic observables can approximate very well the present state of
the system and predictive models based on the dynamics of a reduced
number of macroscopic observables can then be constructed.

In order to identify meaningful low-dimensional structures, we apply
the Proper Orthogonal Decomposition (POD) - also known as the
Principal Component Analysis (PCA) or Karhunen-Lo\`{e}ve expansion -
to an ensemble of surface deformation data generated by the system's
dynamics.  The POD provides the most efficient way of capturing the
dominant components of a dynamical process with only finitely many,
and often surprisingly few, "modes" \cite{hlb}. Using synthetic
calculations for a strike-slip fault system, we show that a reduced
number of deformation modes can explain on the average the large scale
dynamics of elastic surface deformations.  The dynamics of these modes
live on a low-dimensional "reduced" attractor in the neighborhood of
which the system spends most of its time.  The state of the system in
this reduced space - model state space - is represented by a set of
modal coefficients that measures the projection of the ground surface
deformation onto each dominant mode. Our goal is to extract the
nonlinear modal dynamics in this model space by constructing a map of
the observed model state at a future time given the observed model
state at present time.  We will present preliminary results in which
an artificial neural network has been used with promising success to
learn the dynamics of the reduced model from these modal time
series. The reconstructed map has been iterated to obtain predictions
for the long term model dynamics starting from its current model
state. A rough test of the reliability of the model forecast to
approximate the future of the fault system is also discussed.
  
Our method may be compared to the linear pattern dynamics introduced
 by Rundle and his coworkers \cite{rundle:2000}.  Their technique is
 based upon a Karhunen-Lo\`{e}ve expansion of the spatio-temporal
 seismicity data and is used to estimate a linear stochastic model for
 the evolution of a probability density function for seismic activity.
 In contrast to that method, which provides a local linear
 approximation in a probability space, we propose a global nonlinear
 approximation that describes the effective large-scale dynamics in a
 low-dimensional phase space.

\section{2 Description of the Earthquake System}

We distinguish between the assumed physical system and the model which
is an approximate representation of the system. The assumed system is
itself a simplified representation of fault systems in the earth, but
will be considered a valid description of reality if its behavior is
typical for the dynamics of an isolated fault. Our goal is to build
data driven models, with useful predictive skills, which approximate
the large scale dynamics of the system. If this is possible, we can
conclude that the methods we propose can then be extrapolated to real
isolated faults and, perhaps, even to complex fault systems.

\begin{figure}
  \centering\epsfig{file=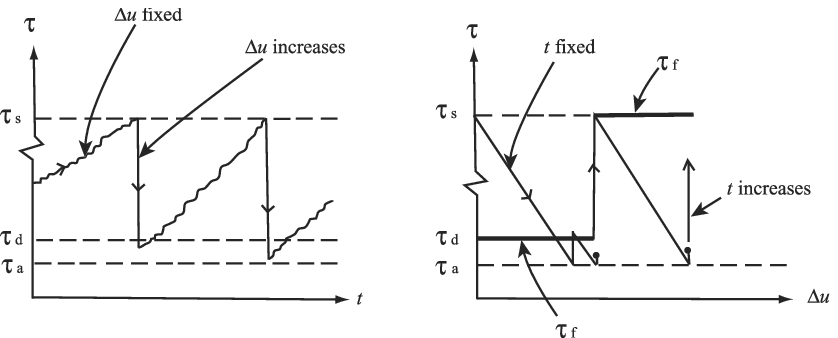,height=50mm,width=150mm}
  \caption{} \label{stress-slip-history}
\end{figure}

Our assumed system corresponds to a discrete strike-slip fault of
length $L=70$ km and width $W=17.5$ km embedded in a $3D$ elastic
continuum \cite{ybz93, ybz96}.  The fault consists of a uniform grid
of dynamical cells where slip is governed by static/kinetic friction
processes, surrounded by regions with imposed constant slip rate of
$V_{pl}=35$ mm/year, representing the tectonic loading.  We divide the
70 km $\times$ 17.5 km computational grid into 128 $\times$ 32 square
cells, with length $\Delta x$ and depth $\Delta z$ having equal
dimensions of approximately 550 m, that corresponds to the dimension of
small, effectively disconnected, slip patches in mature fault zones
\cite{ybz96}.  The brittle deformation at any fault position and time
is governed by quasi-static, $3D$ elastic dislocation theory and
spatially varying ``macroscopic'' constitutive parameters that
describe the static and kinetic friction processes.  The stress at any
fault position (cell) increases with time, $t$ (measured in years),
due to the gradual tectonic loading and the time-dependent brittle
deformation at other fault locations. It can be written as a boundary
integral over the slip deficit  $\phi_{kl} = V_{pl}t-u_{kl}$,
\begin{equation}
\tau_{ij}(t) =\sum_{k,l \in B}F_{ij,kl} \times (V_{pl}t - u_{kl}(t)) \ \ \ ,
\label{stress}
\end{equation}
where $F_{ij,kl}$ is the elastic stress transfer function based on the
solution of Chinnery \cite{chinnery}, $u_{kl}$ is the right lateral
slip at fault cell $(kl)$ ( measured in mm), and the summation is over
the brittle area, $B = $70 km $\times$ 17.5 km, of the fault
plane. The cell indexes $i,k$ and $j,l$ along the horizontal and
vertical directions, respectively, define the spatial cell
location. We assume that the static brittle strength is uniform along
the fault and is set to a value of $\tau_s = 100$ bars.  If the stress
$\tau_{ij}$ reaches the static strength $\tau_{s}$, a brittle failure
occurs at this location and the cell slips an amount $\Delta u_{ij}$,
till the local stress drops to a prescribed arrest stress level,
$\tau_{a,ij}$.  The stress transferred from the failed cell can lead
to subsequent brittle failures (i.e., rupture propagation) if the
stress anywhere increased to the failure threshold. These failures
may, in turn, induce or reinduce more brittle slip events.  After an
initial slip event the strength drops to a dynamic level, $\tau_{d,ij}
< \tau_{s,ij}$, and reinitiation of brittle slip on an already failed
cell occurs when $\tau_{ij} \geq \tau_{d,ij}$ there.  The
 brittle slip associated with a stress drop can be  described by
\begin{equation}
\Delta u_{ij} = \frac{\tau_{f,ij} - \tau_{a,ij}}{F_{ij,ij}} \ \ \ ,
\label{slip_increment}
\end{equation}
where $F_{ij,ij}$ is the self stiffness of a cell and the
failure stress $\tau_{f,ij}$ equals the static failure stress,
$\tau_{s}$, for a cell which has not slipped or is adjusted to its
dynamic value, $\tau_{d,ij}$, every time a cell has failed.  Figure
\ref{stress-slip-history} shows a schematic stress-time and
stress-slip history for an individual cell. During an iteration loop,
all failing cells are identified and their slips updated according to
Eq. (\ref{slip_increment}).  At the end of each iteration the stress
field is updated everywhere according to Eq. (\ref{stress}).  The
iterations end when there are no more brittle instabilities. This
marks the end of an earthquake event whose strength is measured by its
potency $P$ defined \cite{ybz03} as the integral of slip over the
rupture area,
\begin{equation}
P = \sum_{ij} \Delta u_{ij} \Delta A \ \ \ ,
\end{equation}
where $\Delta A = \Delta x \times \Delta z = 0.29$ $\mbox{Km}^2$ and
$\Delta u_{ij}$ is the total slip at cell $(ij)$. During a seismic
event the tectonic slip, $V_{pl} t$, remains constant. At the end of
each event the failure strength everywhere on the fault recovers back
to $\tau_{s}$. In order to initiate the next seismic event we advance
in Eq. (\ref{stress}) the tectonic slip (or equivalently, advance the
time $t$) till the fault cell closest to failure reaches its static
failure strength. This triggers a new seismic event and starts again
the failure iteration loops.

An important component of the fault rheology is the distribution of
arrest and dynamic stresses along the fault. In the simulations used
in this paper we chose random   stresses, 
\begin{equation}
\tau_{a,ij} =
<\tau_{a}> + A \xi \ \ \ ,
\label{arrest_stress} 
\end{equation}
 where $<\tau_{a}>$ is the fault-averaged
arrest stress, $\xi$ is a random number uniformly distributed in the
range $[-0.5,0.5]$, and $A$ is the noise amplitude.  The static
strength, dynamic strength, and arrest stress are related to each
other, at each point on the fault, as
\begin{equation}
D = \frac{\tau_{s}-\tau_{a,ij}}{\tau_{s}-\tau_{d,ij}}   \ \ \ ,
\end{equation}
where $D$ is a dynamic overshoot coefficient. Its inverse, $1/D$, proportional to
$\tau_{s}-\tau_{d,ij}$, is a measure of the dynamic weakening that
characterizes static/kinetic friction models.  The arrest and dynamic
stress distributions do not evolve in time (quenched heterogeneities)
and the fault dynamics is deterministic.  We also assume no
observational errors. We record the fault evolution after all
transients have died out and the dynamics have reached a statistical
steady-state (all moments of the stress and displacement fields become
time independent). A thorough motivation of the fault system, based on
a wealth of observations and numerical results, is provided in a
series of papers by Ben-Zion and Rice \cite{ybz95, ybz93, ybz96},
while a detailed description of the dynamics can be found
in~\cite{ang03, ybz95,zel03}.

\section{3 Qualitative dynamics of slips, stresses, and seismicity}

\begin{figure}
  \centering\epsfig{file=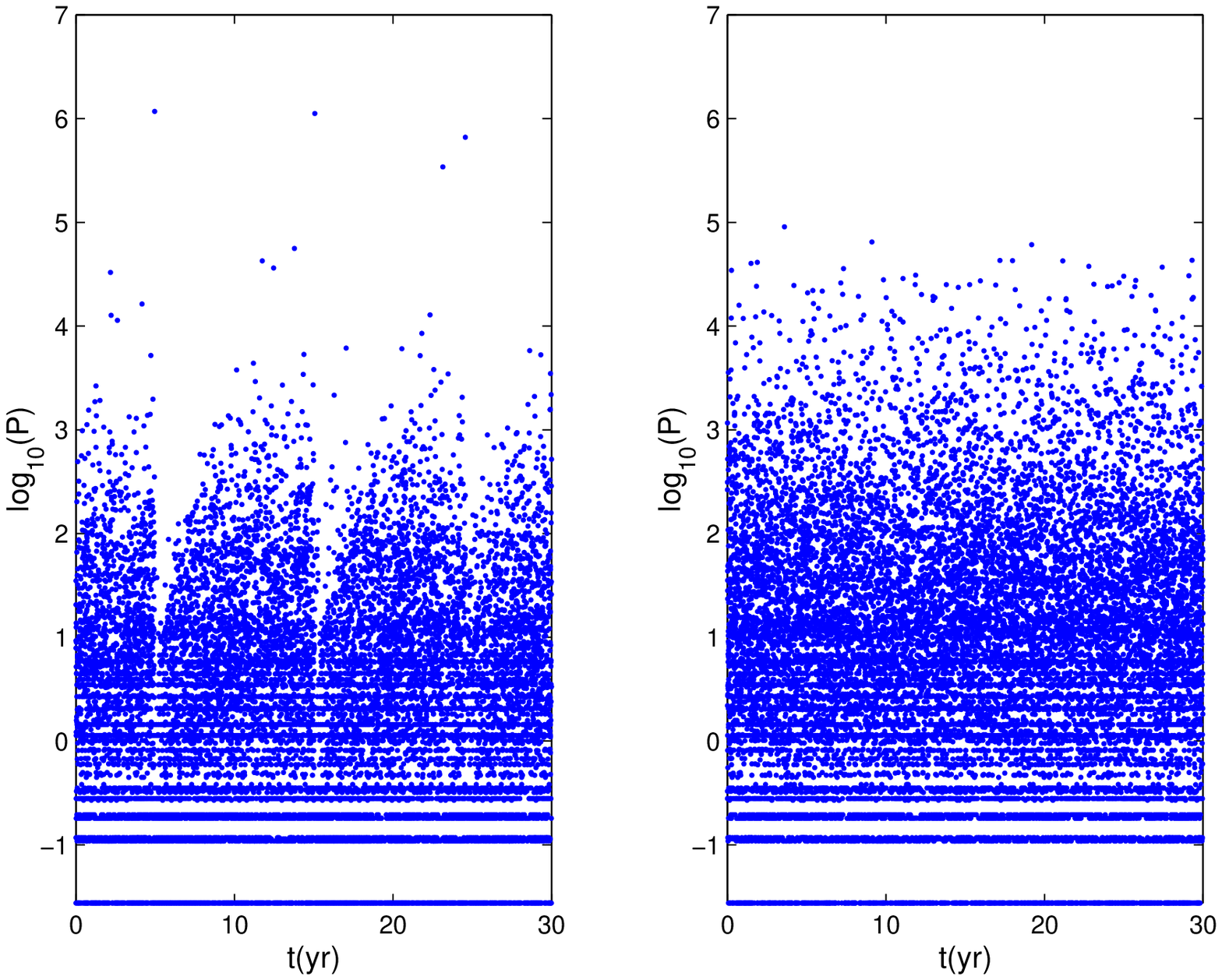,height=50mm,width=150mm}
  \caption{} \label{T_potency}
\end{figure}

\begin{figure}
  \centering\epsfig{file=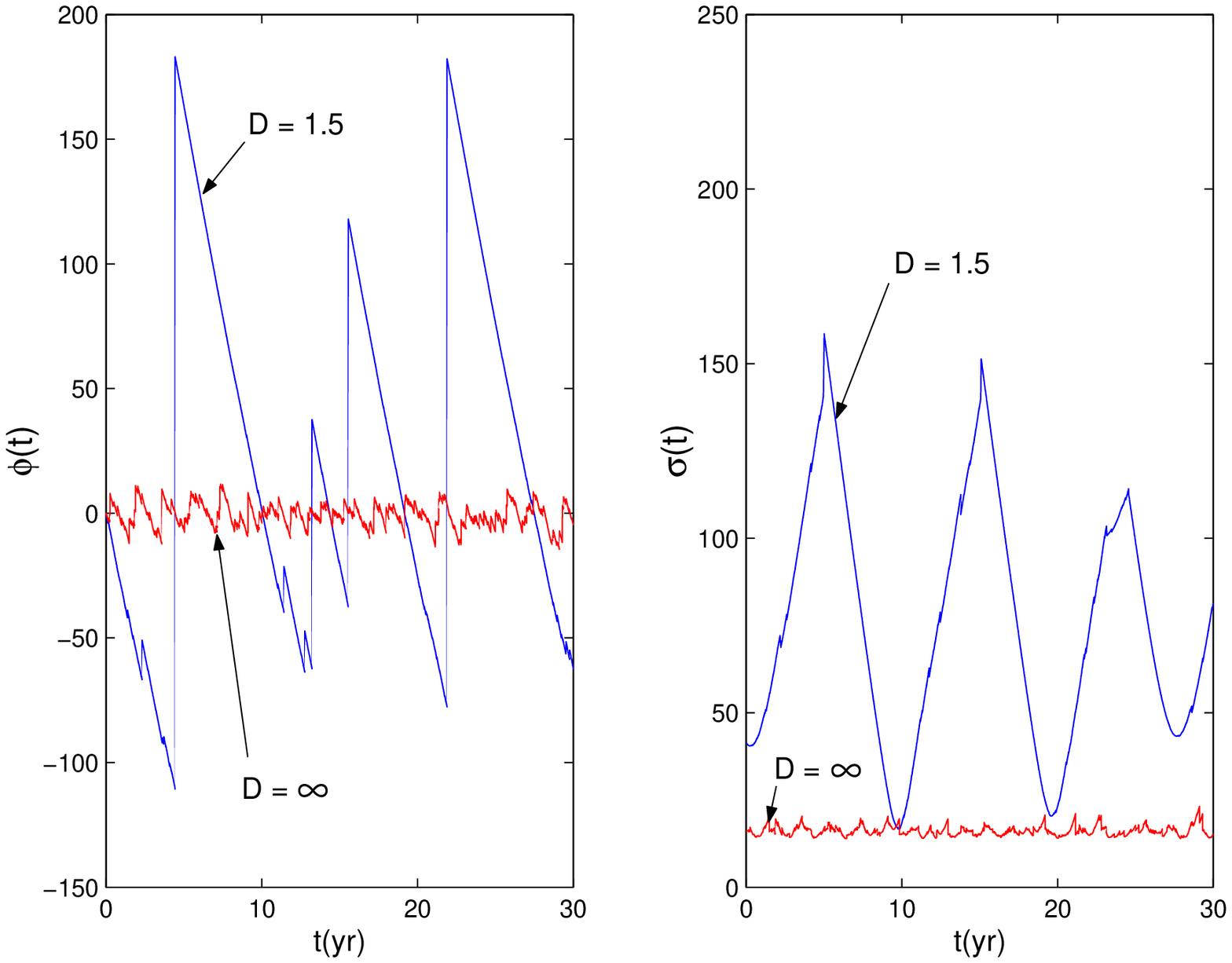,height=50mm,width=150mm}
  \caption{} \label{slip_var}
\end{figure}

We start by analyzing the evolution of seismicity and the dynamics of
the first and second order moments of the stress and slip fields.
Here we only investigate the dynamics of two fault realizations that
differ largely in the value of the dynamic overshoot coefficient.
System $A$ with $D=1.5$ has a large dynamic weakening, while system
$B$ with $D = \infty$ has no dynamic weakening.  
Both systems have random arrest stress distributions, as in
(\ref{arrest_stress}), with $<\tau_{a}> = 80 $ bars and $A=20$ bars
amplitude.  A detailed investigation of the dynamics, covering the
system space between these two limits, will be presented
elsewhere~\cite{ang03}. In general, each system observable contains a
different amount of information for understanding the underlying
dynamics.  As discussed below, for prediction purposes we are only
interested in the dynamics at the large length and time scales.  As
the small scale dynamics overwhelmingly dominates the statistics, it
is not obvious a priori which observable provides information about
the large scale dynamics.  In Fig.~\ref{T_potency} we present the time
evolution of the potency released over a 30 years time interval.  The
fault motion in system B ($D=\infty$) seems to be dominated by
erratic, random appearing evolution. There are no detectable time
patterns.  When significant dynamic weakening is present, as in system
A ($D=1.5$), a clear signature of a quasi-regular evolution emerges.
This can be identified in the presence of very large events (these are
large scale events in which a large fraction of the fault slips),
followed by intervals of seismic quiescence. There is a quasi-regular
build up of correlations~\cite{zel03} which ultimately generates very
large events, much larger in size than in system B. Although the
evolution of seismicity is dominated by small events,
there are
significant evolution patterns reflecting the dynamics of the large
scale events.
These can be better seen in Fig. \ref{slip_var} which shows
 the evolution of the average slip deficit,
\begin{equation}
\phi(t) = \frac{1}{M}\sum_{i,j}(\phi_{ij}(t)-<\phi_{ij}>) \ \ \ ,
\end{equation}
and slip deficit variance,
\begin{equation}
      \sigma(t) = \frac{1}{M}\sum_{i,j}
             (\phi_{ij}(t)-\phi(t))^{2} \ \ \ .
\end{equation}
 Due to the 
inhomogeneity of the slip deficit, the slip fluctuations are measured
from a local time average, $<\phi_{ij}>$, defined as
\begin{equation}
 <\phi_{ij}> = \frac{1}{T}\sum_{t=1}^{T}\phi_{ij}(t) \ \ \ ,
\end{equation}
where the length of the averaging time used is $T=100$ years.
For the $D=1.5$ system we observe large seismic cycles and smooth long
term behavior between two consecutive large events. At the scale of
the plot in Fig. \ref{slip_var}, the erratic and unpredictable small
scale dynamics is unobservable. Magnifying the plot will reveal small
scale jitters superposed over the smooth large scale motion.  This
suggests a clear separation of time scales between the slow, large
scale motions and the fast, small scale motions. In contrast, the
$D=\infty$ system has small oscillations and noisy behavior; note the
large difference in the scale of the variance evolution for these two
systems. For the $D=\infty$ system there are no large scale features
and the dominant signature of small scale events renders the system
higher dimensional.  Additional illustrations of the dramatic change
in the dynamics for cases with and without dynamic weakening are
given in \cite{zel03}.

\begin{figure}
  \centering\epsfig{file=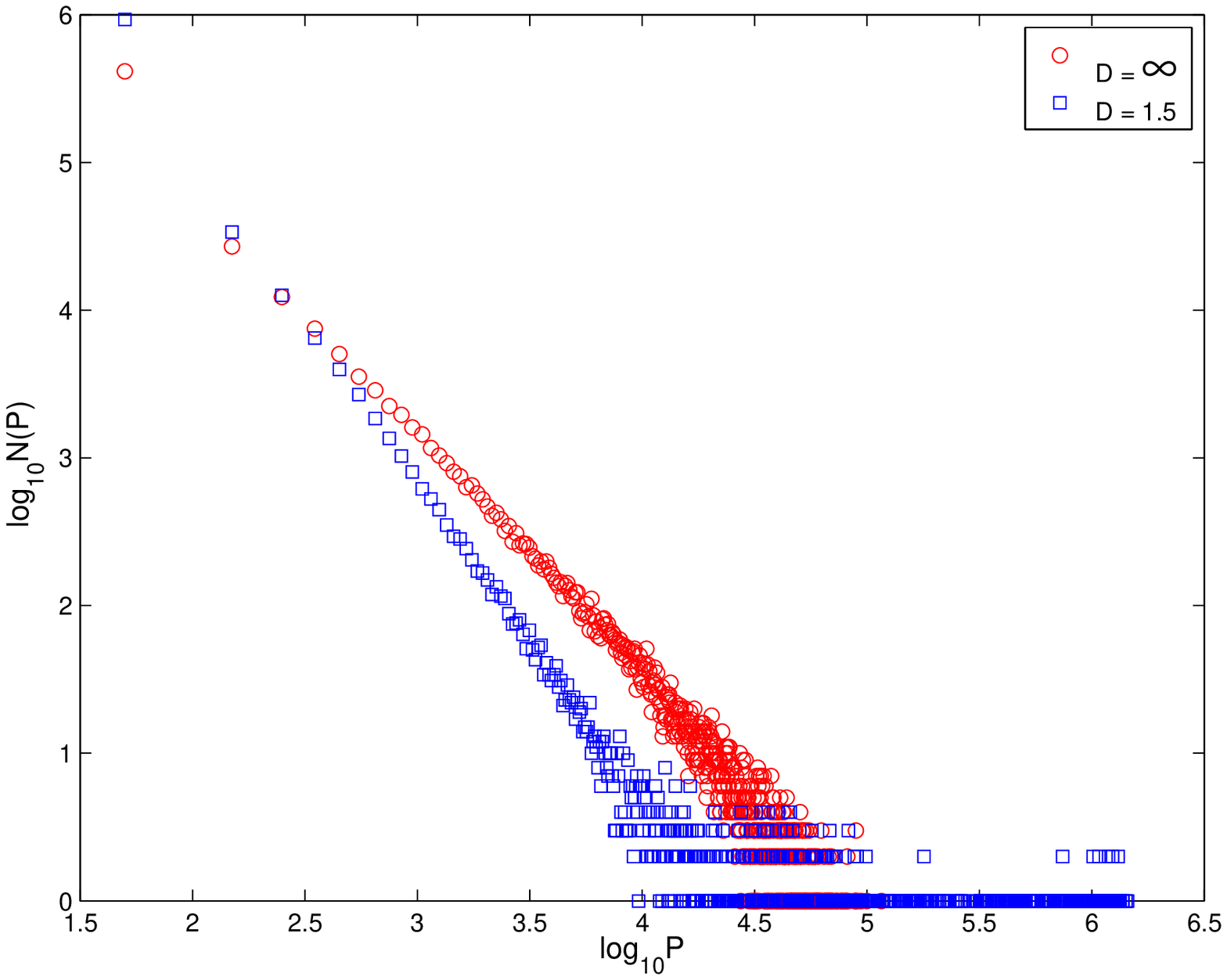,height=60mm,width=60mm}
  \caption{} \label{potency_his}
\end{figure}

The emergence of a clear separation of  length scales is also
revealed in the frequency-size statistics shown in
Fig.\ref{potency_his}. The $D=\infty$ system has a power-law
distribution with an exponential cutoff, characteristic of systems
close to a critical point. This confirms the analysis of Fisher et
al. \cite{fisher97}, who showed that the dynamic weakening
$\tau_{s}-\tau_{d,ij}$ is a tuning parameter and found that the system
has an underlying critical point of second order phase transition at
zero dynamic weakening ($D=\infty$) and full conservation of stress
transfer during failure events.  The frequency-size statistics for the
$D=1.5$ system shows the existence of two separate earthquake
populations. At small scales we observe a power-law distribution of
event sizes. The other events cover a broad range of scales and their
frequency is significantly higher than an extrapolation based on the
power-law statistics of the small events. This behavior is reminiscent
of the characteristic earthquake model~\cite{wes94,ybz03}, which has a
distinct probability density peak at the large size end of the
frequency-size distribution. Moreover, a careful analysis \cite{ang03}
has shown that most stress dissipation is produced at the large size
end of the potency spectrum and is due to the large scale motions.
Averaged over small time intervals, the small events that dominate the
frequency-size histogram do not produce any stress dissipation on
average and only transfer stresses internally.  This suggests a
possible description of the dynamics in terms of large scale
dissipative motions, that balance the stress increment due to an
effective loading rate, which absorbs most of the small scale events
into a renormalized plate velocity. This coarse-grained dynamics is
inherently of lower dimension than the original dynamics. The question
is whether the dimension collapse is significant to render data driven
model reconstruction algorithms practical.


\section{4 Effective dimension of the large scale motions}
\label{dimension}

For the class of systems discussed in this paper the knowledge of the
slip deficit, $\phi_{ij}$, where $i=1,\ldots,N_{x}$ and
$j=1,\ldots,N_{z}$ , fully specifies the state of the fault; the
number of microscopic degrees of freedom is $M = N_{x} \times N_{z}$
and the state of the system is completely represented by a vector in a
finite, but high-dimensional vector space $R^{M}$, where $M=4096$.  We
conjecture that a finite dynamic weakening favors the creation of
spatial and temporal correlations that collapse the asymptotic, long
time dynamics of the fault onto an attractor of a smaller effective
dimension $m$; $m \ll M$.  This is compatible with the recent analysis
of Ben-Zion et al.\ \cite{zel03}. The physics of the dimensional
collapse consists in the clear separation of time and length scales.
In such cases, we can in principle isolate the large scale dynamics
which, evolving slowly, will drive the rapidly relaxing small scale
dynamics.  But, are real faults characterized by a large dynamic
weakening?  Full elastodynamic simulations of rupture propagation in a
3D elastic solid \cite{mad76} indicate that $D \simeq 1.25$,
corresponding to large dynamic weakening. Moreover, we have shown
elsewhere \cite{ang03} that this separation of scales holds for a
broad range of $D$ values and is, therefore, generic. Consequently,
restricting our study to system A, we shall address for the rest of
this paper the following issues (1) Whether we can build a
predictive model for the large scale motions and (2) What effect the
neglected small scale motions have on the predictability of the larger
scales.

In order to model the large scale motions we have to find first if
their dynamics is deterministic in character. In principle, the coarse
grained dynamics of any deterministic system is stochastic. When a
separation of scales exists, we can assume that only the statistical
properties of the small scale motions influence the larger scales,
through coefficients of turbulent viscosity, and that at any moment
these statistical properties are determined by the larger scale
motions \cite{lorenz69}.  A model describing only the large scale
dynamics is assumed to be effectively deterministic.  However, we have
to be careful, because, as pointed out by Lorenz \cite{lorenz69},
there remains uncertainties in these statistics, and their progression
to the very large scales will ultimately produce large scale errors
and limit the forecasting skill of our model.  We will therefore
assume, as a working hypothesis, that a model describing only the
larger scales motions is effectively deterministic. In order to test
our assumption and find an upper bound on the dimension of the model,
we start with a phase space analysis of the large scale motions as is
reflected in the scalar dynamics of the slip deficit variance.

\begin{figure}
  \centering\epsfig{file=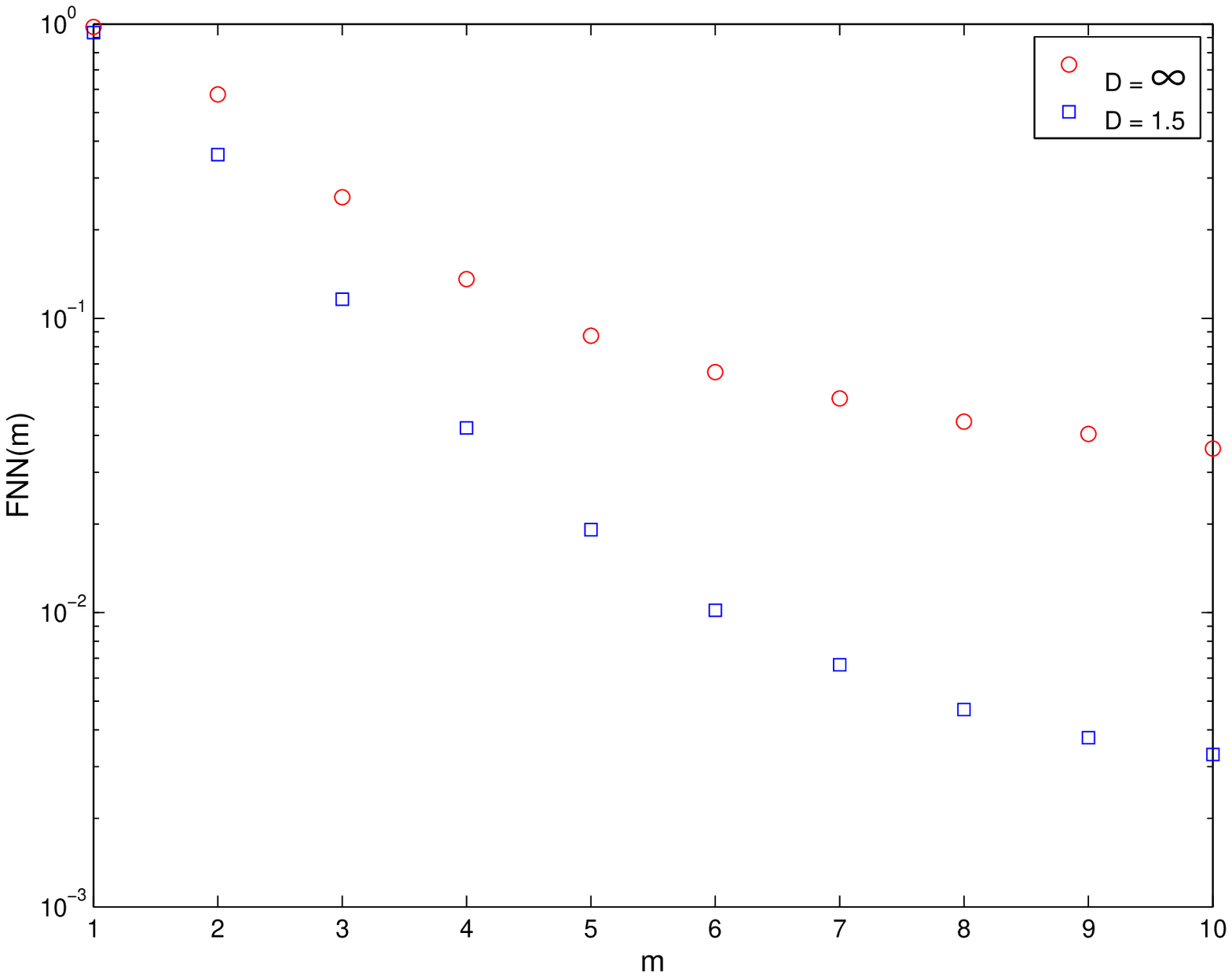,height=60mm,width=60mm}
  \caption{} \label{slip_var_fnn}
\end{figure}

The key to unraveling the dynamics from a scalar time series is to
reconstruct a vector space, formally equivalent to the original
attractor of the dynamics.  For a deterministic system,
Takens \cite{ks97} embedding theorem guarantees that such a
reconstruction is possible from a time-delay reconstruction of only one
scalar observable.  In our case, the embedding is performed using
Cartesian coordinates made out of the observed slip deficit variance
and its time delayed copies,
\begin{equation}
{\bf X}_{n}=[ \sigma(n),\sigma(n-d),\ldots,\sigma(n-(m-1)d)] \ \ \ ,
\end{equation} 
where $\sigma(n)= \sigma(n \Delta t)$ is the slip deficit variance
measured at equal sampling times $\Delta t= 0.1$ years, $d$ is a time
lag (an integer multiple of the common lag $\Delta t$), and $m$ is an
embedding dimension.  The optimal time lag, $d=3$, has been chosen at
the first minimum of the time-delayed mutual information as suggested
in \cite{fs86} and motivated in detail in \cite{ks97}.  The problem is
to determine the optimal embedding dimension $m$, for which the
reconstruction by delay vectors provides an acceptable unfolding of
the attractor so that the orbits composing the attractor are no longer
crossing each other in the reconstructed phase space
\cite{abarbanel:1996}.  The practical solution is to look for false
neighbors in the embedding phase space at a given value of $m$
\cite{ks97}. To understand this concept, consider the situation that
an $m$ dimensional delay reconstruction is an embedding, but an
$(m-1)$ dimensional delay reconstruction is not.  If the embedding
dimension is too small to unfold the attractor, a small $R^{m-1}$
neighborhood will contain points that belong to different parts of the
original attractor.  Therefore, at a later time, the images of these
points under the system's dynamics will split onto different groups,
depending on which part of the attractor the points are originally
coming from. This lack of a unique location of all the images in
$(m-1)$ dimensions is reflected in finding false neighbors, meaning
that determinism is violated. When increasing $m$, starting with small
values, one can detect the minimal embedding dimension, $m_e$, by
finding no more false neighbors. Then, we can invoke the result of
Sauer et al. \cite{sauer91} who showed that the attractor formed by
${\bf X}_{n}$ in the embedding space is equivalent to the attractor in
the unknown space in which the system is living if $m$ is larger than
twice the box counting dimension, $d_{0}$, of the attractor. Often, an
$m=d_{0}$ embedding dimension is enough for unfolding the
attractor. Therefore, the minimal embedding dimension sets the
following bounds, $(m_e-1)/2 \le d_0 \le m_e$, for the dimension of
the attractor in the true phase space.  Figure \ref{slip_var_fnn}
shows that the percentage of false neighbors for the $D=1.5$ system
falls under $ 1\%$ when the phase space embedding of the slip variance
data reaches dimension seven.  This behavior suggests a topological
dimension in the true phase space between three and six. We note that
the percentage of false neighbors is not completely reduced to
zero. The explanation for this behavior is discussed in Section 6.  To
appreciate this result, we show for comparison in
Fig. \ref{slip_var_fnn} the behavior of the system in the limit of no
dynamic weakening. Even for an embedding dimension as high as ten, the
percentage of false neighbors does not drop below $4 \%$, and is an
order of magnitude larger than the percentage for the $D=1.5$ system
at the same embedding dimension. This confirms our previous analysis
and proves that in the limit of low dynamic weakening the system has
high dimensional dynamics.


The foregoing analysis shows convincingly the deterministic structure
of the large scale motions: for stochastic or very high dimensional
dynamics the number of false neighbors will never effectively drop to
zero. The low-dimension result is consistent with the embedding
theorem \cite{sauer91} which asserts that it is not the dimension of
the underlying true phase space that is important for the minimal
dimension of the embedding space, but only the fractal dimension of
the support of the invariant measure generated by the dynamics in the
true phase space \cite{ks97}. These results justify our earlier
remarks that due to some collective behavior only a few dominant
degrees of freedom remain and our conjecture that their dynamics is
effectively deterministic.  

\begin{figure}
  \centering\epsfig{file=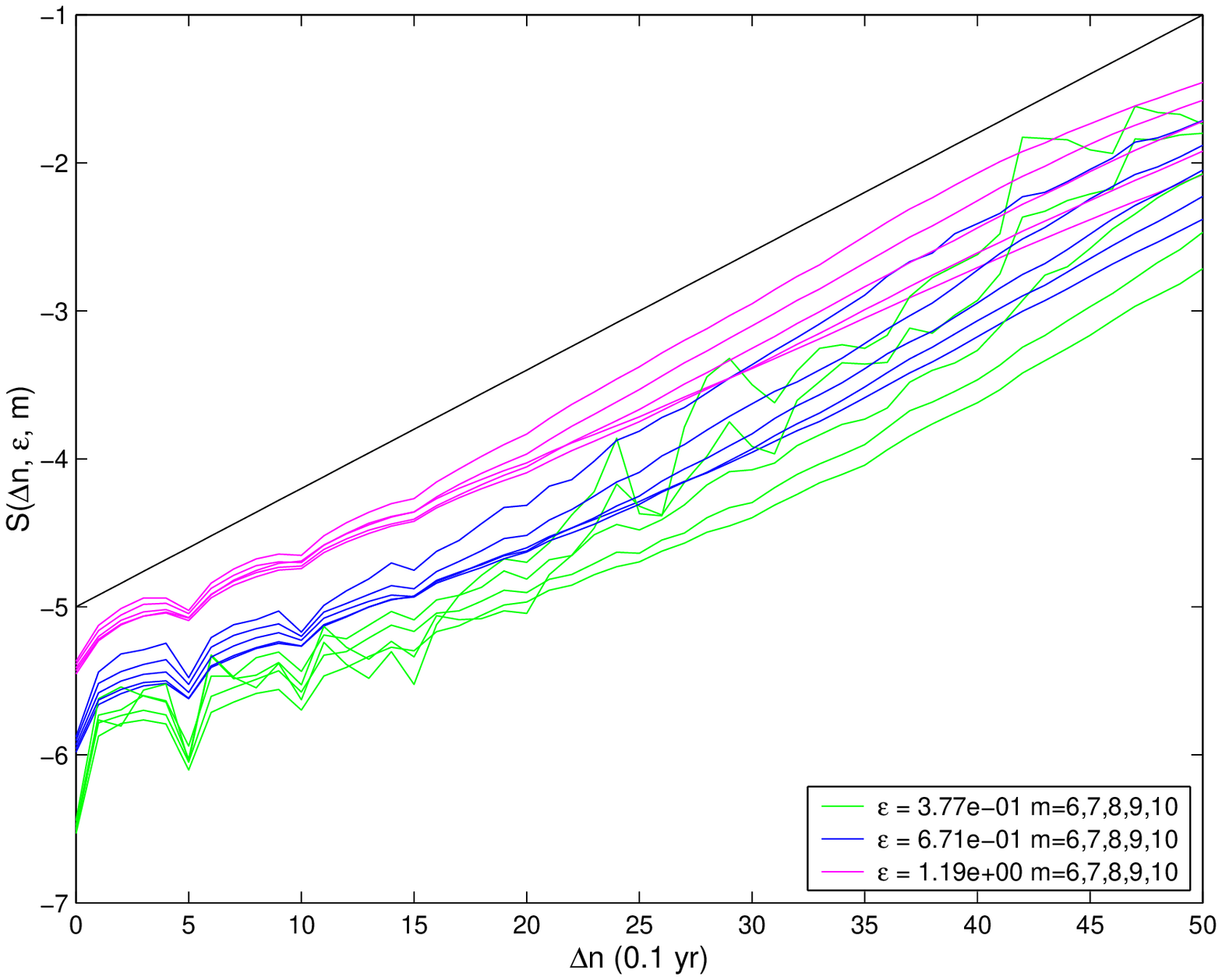,height=60mm,width=60mm}
  \caption{} \label{slip_var_lyap}
\end{figure}

Before addressing the problem of identifying these collective degrees
of freedom, we would like to know if their dynamics is chaotic and, if
so, what are the implications for their predictability. To answer this
question we compute the maximal Lyapunov exponent.  This exponent
measures the exponential divergences of nearby trajectories and is an
average of these local divergences over the whole data.  A positive
maximal Lyapunov exponent is a signature of chaos.  Its computation is
based on the algorithm described by Kantz and Schreiber \cite{ks97}
and uses the implementation provided in the publicly available
software package TISEAN \cite{hks99}(all nonlinear time series
algorithms described in this paper use the same excellent
implementation).  For a point ${\bf X}_{n}$ of the time series in the
embedding space, the algorithm determines all neighbors ${\bf X}_{n'}$
within a neighborhood $U_n$ of radius $\epsilon$. Then, for each
neighbor, we consider the distance $\delta(0) = |{\bf X}_{n} - {\bf
X}_{n'}|$ and we read its evolution $\delta(\Delta n) = |{\bf
X}_{n+\Delta n} - {\bf X}_{n'+ \Delta n}|$ from the time series. If
for all neighbors $\delta(\Delta n) \simeq \delta(0) e^{\lambda \Delta
n}$, then $\lambda$ is the local divergence rate.  The local
divergence rate varies along the attractor and the true maximal
Lyapunov exponent is an average over many reference points $n$.
Therefore, in practice we compute the average over the distances of
all neighbors to the reference part of the trajectory as a function of
the relative time $\Delta n$. The logarithm of the average distance at
time $\Delta n$ measures the effective expansion rate over the time
span $\Delta n$:
\begin{equation}
   S(\Delta n,\epsilon,m) = \langle \ln (\frac{1}{| U_{n}|}
     \sum_{{\bf X}_{n'} \in U_{n}} 
     |{\bf X}_{n+\Delta n}-{\bf X}_{n'+ \Delta
     n}|)
   \rangle_{n} \ \ \ ,
\end{equation}
where the outer bracket denotes the averaging over the reference point
${\bf X}_{n}$, $| U_{n}|$ is the number of neighbors in $U_n$, and the
argument of the logarithm is the average local expansion rate.  If
$S(\Delta n,\epsilon,m)$ vs $\Delta n$ exhibits a robust linear
increase with identical slope for a range of $\epsilon$ values and for
all $m$ larger than some minimum embedding value, then its slope is an
estimate of the maximal Lyapunov exponent $\lambda_{max}$ per time
step \cite{ks97}. Figure~\ref{slip_var_lyap} shows three bundles of
curves corresponding to three neighborhood sizes, $\epsilon = 0.377,$
$0.671$, and $1.19$, as can be seen for $\Delta n = 0$. Each bundle
shows the behavior of the expansion rate for $m=$ 6,7,8,9 and 10 and
proves that the result is robust to changes in $\epsilon$ and does not
depend on the embedding dimension when $m$ is large enough. Our
estimate for the maximal Lyapunov exponent is $\lambda_{max} = 0.8
\mbox{\,\,years}^{-1}$.  We can therefore conclude that the large scale
motions
have a positive maximal Lyapunov exponent and exhibit sensitive
dependence on the initial conditions. Does this dependence sets a
fundamental limitation to long-term forecasting?  Two initially close
trajectories will diverge exponentially in the phase space with a rate
given by the largest Lyapunov exponent $\lambda_{max} $. Therefore,
for any finite uncertainty $\delta$ in the initial conditions, we can
forecast the future state of the system only up to a maximum time,
\begin{equation}
T_p \simeq \frac{1}{\lambda_{max}}\ln (\frac{\Delta}{\delta})
\end{equation}
where $\Delta$ is the accepted tolerance level. For the earthquake
system, an approximate estimate for the predictability time is $T_{p}
= 1/\lambda_{max} \simeq 1.25 $ years.  This result is disappointing
and, due to expected model reconstruction errors, which might
overshadow the uncertainty in the initial conditions, raises questions
on the value of any forecasting endeavor.  But this estimate is naive
and holds only for infinitesimal perturbations and in nonintermittent
systems.  Generally, the predictability time is scale dependent
\cite{bcfv02} and can be much longer than the rough estimation $T_{p}
\simeq 1/\lambda_{max}$.  As we will shortly see, at the large length
scale of interest for forecasting, a better estimation of the
predictability time is $T_{p} \simeq 10$ years.

\section{5 Proper Orthogonal Decomposition}
\label{POD}

The dynamics of the first and second order moments of the slip and
stress fault fields provides an adequate description of the large scale
motions. Unfortunately, their measurement poses very difficult
problems: the stress field is not directly observable while the slip
field requires the solution of an ill-posed inverse problem.
Therefore, we propose an alternative analysis of the large scale
motions based on the spatio-temporal strain patterns embedded in the
surface deformation fields that can be accurately measured by InSAR
and GPS observations \cite{brf00}.
  Assuming uniform slip discontinuities over
 each rectangular cell, the components of the surface displacement vector
 $U_{\alpha}$ are found to be \cite{okada},
\begin{equation} 
U_{\alpha}({\bf x}) =
\sum_{i,j}\Gamma^{\alpha}({\bf x};{\bf x}_{ij}) 
\times (u_{ij} - V_{pl}t) \ \ \ , 
\label{surf_disp}
\end{equation}
where $\Gamma^{\alpha}({\bf x};{\bf x}_{ij})$ is the surface
displacement in the direction $\alpha = x,y,z$ at the observation
point ${\bf x}=(x,y,0)$ due to uniform unit slip on the fault cell
$(ij)$ located at ${\bf x}_{ij}= (x_{ij},0,z_{ij})$. The space-time
signal is obtained by simultaneous measurements of the surface
displacements on a $64 \times 32$ uniform rectangular grid which
covers a $100 \mbox{ km} \times 50 \mbox{ km}$ surface area centered
around the fault.  For each surface deformation we compute an ensemble
of $N$ snapshots, $ {U_{\alpha}({\bf x},n)} = {U_{\alpha}({\bf x}, n
\Delta t)},$ $ n=1,\ldots,N,$ and $\alpha = x,y,z$ , every $\Delta t =
0.1$ years.
The analysis of each
deformation direction proceeds identically, therefore we will
henceforth drop the deformation index $\alpha$. Since we are only
interested in decomposing the dynamics of fluctuations, it is
convenient to separate the flow $U({\bf x},n)$ into a time-independent
mean and a fluctuating part, i.e.
\begin{equation}
    U({\bf x},n)= <U({\bf x})> +
    u({\bf x},n) \ \ \ ,
\end{equation}
where $<u({\bf x},n)> = 0$ and $< \cdot > $ indicates the ensemble
average. 

The identification of the active degrees of freedom in the surface
deformation fields uses the Principal Orthogonal Decomposition (POD)
\cite{hlb}. 
The POD seeks an optimal representation (in the least square sense) for
the members of the ensemble $\{u({\bf x},n)\}_{n=1}^{N}$. It
searches for generalized directions $\phi({\bf x})$ in the
configuration phase space, such that most of the ensemble fluctuations
will be directed along $\phi({\bf x})$. Mathematically, this is
achieved by maximizing the average projection of the $\{u({\bf
x},n)\}$ ensemble onto $\phi({\bf x})$, i.e.
\begin{equation}
\underset{\phi \in L^2, \| \phi \|= 1}{\max} < (u,\phi)^2 > \ \ \ ,
\end{equation}
where $( \cdot , \cdot )$ is the $L^2$ inner-product in the
configuration space and the $L^2$ norm constraint, $\| \phi \|= 1$, is
required for the maximum to be defined.  The solution to this
variational problem is given by the eigenfunctions $\{\phi_{i}
\}_{1}^{M}$ of the following integral equation \cite{hlb},
\begin{equation}
\int K({\bf x},{\bf y}) \phi_{i}({\bf y})d{\bf y} =
\lambda_{i} \phi_{i}({\bf x}) \ \ \ , \hspace{0.5cm}
\lambda_{1} \ge \lambda_{2} \ge \ldots \ge \lambda_{M} \ge 0 \ \ \ , 
\end{equation}
whose kernel $K({\bf x},{\bf y})$ is the two point correlation
matrix of the ensemble, $K({\bf x},{\bf y}) = < u({\bf x}) u({\bf
y}) > $.
There are at most $M$ eigenfunctions corresponding to the total number
of degrees of freedom.  The eigenvalues $\lambda_{i}$ measure the mean
square fluctuations of the ensemble in the directions defined by their
corresponding eigenfunctions: $\lambda_{i} = < (u,\phi_{i})^2 >$.
We can also think of $\lambda_{i}$ as the average ``energy'' of the
ensemble fluctuations projected onto the $\phi_{i}$ axis.  Therefore,
ranked in decreasing order of their eigenvalues, the eigenfunctions
(sometimes called empirical eigenfunctions, coherent structures, or
dominant modes) will identify the dominant directions in configuration
space along which most of the fluctuations take place.  The first four
modes describing the predominant motions of the $u_{x}$ deformation
field for system A with $D=1.5$ are shown in Fig. \ref{ux_evecs}. The
coherent, collective nature of the fluctuations represented by these
modes is eloquently expressed by their spatial structure. The spatial
coherence sharply decreases for the higher order modes (not shown).
The modes probe the system at different scales, from the largest
scales, captured by the most dominant modes, to the smallest scales,
described by the higher order modes. This observation is hardly
surprising, since for a translationally invariant system the POD modes
are simply the Fourier modes \cite{hlb}.

The basis $\{ \phi_i \}_{i=1}^{M}$ is optimal in the sense that
any reduced representation of the form
\begin{equation}
u({\bf x},n) \simeq 
\sum_{i=1}^{m} A_{i}(n) \phi_{i}({\bf x}), \hspace{0.5cm} m < M \ \ \ ,
\label{PODembedding}
\end{equation}
describes typical members of the ensemble better than {\it any} linear
representation of the same dimension $m$ in any other basis: the
leading $m$ POD modes contain the greatest possible ``energy''
on average \cite{hlb}. Therefore, due to the coherence and optimality
of the POD basis,
 Eq. (\ref{PODembedding}) provides the most efficient Euclidean
 (linear) embedding for the large scale motions of the system.  
 Moreover, the modal coefficients are uncorrelated on
 average, i.e.
\begin{equation}
<A_{i} A_{j} > = \lambda_{i} \delta_{ij} \ \ \ ,
\label{mode_independence}
\end{equation}
which reflects the fact that orthogonality in the
embedding space, $(\phi_{i}, \phi_{j} ) = \delta_{ij}$, is related to
the statistical properties of the time series. 

The choice of the embedding dimension $m$ is based on the computation
 of the cumulative normalized eigenvalue spectrum, $\Lambda_{m}$,
 defined as:
\begin{equation}
\Lambda_{m} = 
\frac{\sum_{j=1}^{m}\lambda_{j} }
       {\sum_{j = 1}^{M} \lambda_{j} } \ \ \ .
\end{equation}
The cumulative spectrum can help us define an effective POD embedding
dimension by finding the minimum number of modes needed to
capture some specific fraction $f<1$ of the total variance of the
data:
\begin{equation}
d_{POD} = \operatornamewithlimits{arg\, min}_{m}
        \{\Lambda_{m} : \Lambda_{m} > f\} \ \ \ .
\end{equation}
Figure ~\ref{cum_evals} shows the cumulative normalized spectrum for
each of the surface deformation modes.  For system A ($D=1.5$), in
order to explain $f=95\%$ of the variance in the $u_{x},u_{y},u_{z}$
data sets we need only 2, 3, and 5 modes respectively. This is
consistent with our earlier estimates of the embedding dimension. In
contrast, for system B ($D=\infty$), we need 9, 27, and 42 modes
respectively.

It is important to realize that the POD modes carry no dynamic
information.  Reshuffling the ensemble of configurations - and,
therefore, destroying the dynamic content hidden in their time
ordering - produces the same POD modes.  Therefore, in order to
identify the dynamics of each POD mode we calculate the projections of
the surface deformation fields onto the dominant deformation modes,
i.e.,
\begin{equation}
  A_{i}(n) = (u({\bf x},n), \phi_{i}({\bf x}) ) \ \ \ .
\end{equation}
This way, we generate a small number of modal time series $A_{i}(n),
i=1, \ldots,m,$ that encode the evolution, interaction and dynamics of
the spatial modes \cite{hlb}.  They encapsulate the projection of the
system's dynamics onto the $m$-dimensional {\it model} space defined
by Eq. (\ref{PODembedding}).

In Fig. ~\ref{ux_amp_evol} we represent the modal time series (red
lines) corresponding to the first four $u_{x}$ spatial modes for fault
system A. The blue lines describe the time evolution of the binned
potency released (the total potency released over the time interval
$\Delta t$). For the first time series, $A_{1}(n)$, the location in
time of the amplitude jumps coincides with the time of large events
(within the temporal resolution defined by $\Delta t$).  Moreover, we
can show that the size of the amplitude jumps is proportional with the
size of the event, i.e. $\Delta A_{1}(n)=$ $ A_{1}(n+1) - A_{1}(n)$ $
\simeq P_{max}(n)$, where $P_{max}(n)$ is the largest event in the
$n$'th time interval. Therefore, if we can model the evolution of this
mode, the time and size of the amplitude jumps will give us useful
information about the time and size of the large earthquake
events. Moreover, we notice that the higher order modes evolve on
faster time scales, as is already noticeable from the faster decay of
the temporal autocorrelations shown in Fig. \ref{time_corr}.

The picture that
emerges from this decomposition evolves gradually from the slow,
coherent large scale motions, whose dynamics is captured by the most
dominant POD modes, to the fast, incoherent small scale motions
described by the higher order modes. In Section 7
\ref{model_reconstruction} we  describe how a neural net can be
used to process these time series in order to extract a
low-dimensional nonlinear dynamic model with short-term predictive
capabilities.




\section{6 Geometry of the attractor in the POD basis}
\label{PODattractor}

The POD modes provide an optimal basis for the phase space embedding
expressed in Eq. (\ref{PODembedding}).  The phase space vectors in
this reconstruction, ${\bf X}_n =(A_{1}(n),\ldots,A_{m}(n))$,
approximate the location of the system on the attractor at discrete
time moments. Compared to the time-delay embedding, the POD basis
unfolds the attractor with increased resolution as the embedding
dimension increases.  This property allows us to probe the structure
of the attractor at different length scales by computing its
correlation dimension.  This measure was introduced by Grassberger and
Procacia \cite{gp83} to quantify the self-similarity of geometrical
objects. It is based on the definition of the correlation sum for a
collection of points ${\bf X}_{n}$ in some vector space $R^{m}$.  This
is the fraction of all possible pairs of points which are closer than
a given $\epsilon$ in a particular norm. The basic formula is
\begin{equation}
C(m,\epsilon) =
\frac{2}{R(N-2W-1)}\sum_{i=1}^{R}\sum_{j=1,|i-j|>W}^{N}
   \Theta(\epsilon-  \parallel {\bf X}_{i}-{\bf
     X}_{j}\parallel_{\infty} ) \ \ \ ,
\end{equation}
where we use the maximum norm, $\parallel {\bf X} \parallel = \max
|x_{i}|$, $\Theta$ is the Heaviside step function, $N$ is the total
number of data points, and $R$ is the total number of reference
points. The sum counts the pairs $({\bf X}_{i},{\bf X}_{j})$ whose
distance is smaller than $\epsilon$.  Note that data points within a
time window $2W$ of any reference point $j$ are not included in the
correlation sum.  An appropriate choice of $W=100$ reduces the spoiling
effect of autocorrelations from the correlation sums as suggested by
Theiler \cite{th90}. In the limit of an infinite amount of data and
small $\epsilon$, the attractors of deterministic systems show power law
scaling, i.\ e., $C(\epsilon) \simeq \epsilon^{d_2}$, and we can define
the correlation dimension $d_2$ by
\begin{equation}
d_{2} = \underset{\epsilon \rightarrow 0}{\lim}
        \underset{N \rightarrow \infty} {\lim}
  d_{2}(m,\epsilon) \ \ \ .
\end{equation}
For each embedding dimension $m$, $d_{2}(m,\epsilon)$ measures
the local slopes of the correlation sum at different length scales
$\epsilon$,
\begin{equation}
d_{2}(m,\epsilon) = \frac{\partial \ln C(m,\epsilon)}
                  {\partial \ln \epsilon} \ \ \ .
\end{equation}
In Fig. \ref{d2}. for system A on the left and B on the right, we show
the $\epsilon$ dependence of $d_{2}(m,\epsilon)$ for embedding
dimensions $m=1,\ldots,20$. This plot allows the identification of a
scaling range and estimation of the correlation dimension if such
range occurs.  This is reflected in the presence of a plateau in the
$\epsilon$ dependence of $d_{2}(m,\epsilon)$ which does not change
much with embedding dimension $m$ when $m > d_{0}$: the correlation
sum probes the attractor at different length scale and tests for
scale-invariance.

 From the point of view of the physics involved, we distinguish four
different types of behavior for $d_{2}(m,\epsilon)$ at different
regions of length scale \cite{ks97}.  For small $\epsilon$ and large
$m$ (region I in Fig.~\ref{d2}) the lack of data points is the
dominant feature and the values of $d_{2}(m,\epsilon)$ are subject to
large statistical fluctuations. If $\epsilon$ is of the order of the
size of the entire attractor (region IV), no scale invariance can be
expected.  In between, we can distinguish two regions. In region II,
at small length scales but good statistics due to the small embedding
dimension, the reconstructed points reflect the low amplitude and high
frequency dynamics generated by the small events.  At these length
scales the dynamics is high dimensional and the reconstructed points
fill the entire phase space available, therefore we expect
$d_{2}(m,\epsilon) \simeq m$. Up to the embedding dimension $m=4$ this
estimate is recovered in the limit $\epsilon \rightarrow 0.$ For
increased embedding dimension there are large statistical fluctuations
due to the lack of neighbors.

For system A, in region III located at larger length scales between
regions II and IV, the dynamics evolves on a low-dimensional
self-similar attractor whose presence is detected by the plateau in
the correlation dimension at $d_{2}(m,\epsilon) \simeq 1.3$.  Note
that for system A the scaling regime, $\epsilon \rightarrow 0$, is not
reachable. The breakdown of scaling at the smaller length scales is
dynamic in nature and is not due to the lack of good statistics. The
closer we look at the system, the more degrees of freedom become
visible and the dimensionality of the dynamics is higher than our
largest embedding dimension. This also explains why the percentage of
false nearest neighbors never drops completely to zero with increased
embedding dimension. On the same figure, within the range of embedding
dimensions that were numerically accessible, system B shows no
signature of a large scale dimension collapse, in dramatic contrast
with the behavior of system A.

 The length dependent dimensions lead us to conjecture that we observe
different subsystems on different length scales. From the self-similar
and low-dimensional structure at the large length scales the structure
of system A  attractor crosses-over to a high dimensional, stochastic like
structure when probed at very small length scales.  Our strategy for
building a model with good forecasting skills should be obvious
now. The underlying idea is to approximate the large scale motions of
 system A through a low-dimensional embedding into the subspace
spanned by the most dominant POD modes and to extract a
finite-dimensional model in the form of a set of ODEs of comparable
dimension. Does a good nonlinear model for the large scale motions
exists?  The answer depends on how rapidly the small scale motions of
the {\it true} system progress to reach the larger length scales.  If
this time is short, an effective model for the large scale motions
might not exist. If the growth time is sufficiently long, the
existence question is well posed and a model for the large scale
motions might exist. To address this question we  now
proceed to the model reconstruction task.

\section{7 Model Reconstruction and Short-Term Earthquake Forecasting}
\label{model_reconstruction}

Using the POD decomposition we have identified a low-dimensional
linear space in which system A evolves most of the time and we have
reduced its dynamics to a small set of time series, $A_{i}(n)$,
$i=1,\ldots,m$, describing the evolution of the system in this reduced
linear space. We now face the problem of determining the underlying
dynamical process from the information available in these time series.
They are assumed to be governed by a nonlinear set of ODEs and our
modeling approach relies on the ability to identify an approximate $m$
dimensional model,
\begin{equation}
A_{i}(n+1) = F_{i}[A_{1}(n),A_{2}(n),\ldots,A_{m}(n)] 
\hspace{0.3cm}i=1,\ldots,m \ \ \ ,
\label{minimal_model}
\end{equation}
that describes an explicit Euler approximation to the evolution and
interaction of the  spatial modes.  To identify this
nonlinear mapping we employ an artificial neural network (ANN).
Generally, 
the neural network approach is used as a ``black-box'' tool in order
to develop a dynamic model based only on observations of the system's
input-output behavior \cite{rkk92,rkk95}.  In the learning process the
network adjusts its internal parameters to minimize the squared error
between the network output and the desired outputs.  A typical
learning method is the error back-propagation algorithm which is a
first order gradient descent method~\cite{haykin}.

All reconstruction results described here refer to the dynamics of the
$u_x$ surface deformation modes for model A with $D=1.5$.  We found it
difficult to identify a simple model having the structure defined by
Eq.~(\ref{minimal_model}). In order to improve the model forecasting
skill it is useful to enlarge the structure of the model to include
information about the past history of the modes. While the best model
structure is still a subject of investigation, an analysis of the time
patterns present in the modal evolution provides a partial
understanding of this result.
As Fig. \ref{ux_amp_evol} shows, one essential feature of the modal
dynamics is the presence of two time scales: within each earthquake
cycle there are intervals of slow and fast motions with  detectable
quasi-regular behavior. This is  reflected in the two-point autocorrelation
functions (Fig. \ref{time_corr}) defined as
\begin{equation}
C_{ii}(\tau) = \frac{<A_i(n ) A_i(n+\tau)> - <A_i(n)>^2}{<A_i(n)>^2} \ \ \ , 
\end{equation}
where the $<\cdot>$ denotes the time average.
The resulting plots exhibit oscillatory behavior with a slow amplitude
decay over a longer time scale indicating that the system has two
correlation time scales.
We have observed that providing the ANN with information about these
long-term correlations of the modal dynamics produces nonlinear
models with increased forecasting skill. To include this information,
we have modified the structure of the model in
Eq. (\ref{minimal_model}) and replaced the modal coefficients $A_{i}$
with time-delayed vectors ${\bf X}_{i}$,
\begin{equation}
{\bf X}_{i}=( \underset{\mbox{short-time memory}}
           {\underbrace{A_{i}(n),A_{i}(n-1),\ldots,A_{i}(n-d)},}
         \underset{\mbox{long-time memory}} 
   { \underbrace{A_{i}(n-2d),\ldots,A_{i}(n-(K-1)d)}} ) \ \ \ ,
\label{mode_vector}
\end{equation}
 where $d$ is the delay and $(K-1)$ is the number of the time-delay
intervals.  Due to the presence in the modal dynamics of two different
time scales, we have included both short,
$A_{i}(n),A_{i}(n-1),\ldots,A_{i}(n-d)$, as well as long time,
$A_{i}(n-2d),\ldots,A_{i}(n-(K-1)d)$, memory information in
the time-delayed input vectors.  The dimension of each time-delayed
vector is $K+(d-1)$.  The ANN output then provides a prediction of the
mode amplitude $A_{i}$ at time $(n+1)$,
\begin{equation}
A_{i}(n+1) = {F}_{i}[{\bf X}_{1}(n), {\bf X}_{2}(n), 
\ldots,{\bf X}_{m}(n)]\hspace{0.3cm}i=1,\ldots,m \ \ \ ,
\label{long_model}
\end{equation}
based on input information describing the past mode histories.

The ANN is trained using a standard back-propagation algorithm
\cite{haykin}. When training succeeds, the ANN provides an approximate
dynamical model (map) for the large scale motions of the fault. The map can
be evaluated once to provide short term predictions or iterated to
predict the long term fault dynamics.  Besides the parameters
describing the structure of the ANN (nonlinear transfer functions,
number of layers and neurons in each layer), the model structure
itself has many parameters ($m$, $d$, $K$) that can be adjusted in
order to improve the model performance.  Our goal is to find a good
model describing the evolution of the first POD mode, whose dynamic
discontinuities trace accurately the time and size of the large
events. Information about the higher order modes and their past
histories is only necessary to uniquely determine the state and 
 model trajectory in its phase space.  One of the best models found so far
describes the evolution of the first two $u_{x}$ surface
deformation modes: $m=2$. For each spatial mode, a time-delayed vector
with parameters $K=6$ and $d=6$ was used.  Because the dimensionality
of the input space, $m(K+(d-1)) = 22$, is higher than our estimated
embedding dimension, we have decided to perform a POD analysis of the ensemble
of input vectors. This time, the POD decomposition performs the
analysis of the dominant {\it temporal} patterns that are created by the
modal dynamics. Similar to the spatial decomposition, we first
compute the two point correlation matrix of the ensemble,
\begin{equation}
K_{ir,js} = < { X}_{ir} { X}_{js} > \ \ \ ,
\end{equation}
where $i,j = 1, \ldots,m$ denote the input modes, $r,s = 1, \ldots,
K+(d-1)$ are the components of the time-delayed mode vectors, and $<
\cdot >$ is the average over the ensemble of input vectors. Due to the
statistical independence of the modal coefficients,
Eq. (\ref{mode_independence}), the correlation matrix is to a good
approximation block diagonal, i.e. $K_{ir,js} \simeq 0 $ for $ i \ne
j$ (any modal cross-correlations are due to finite size effects).  The
eigenvalues and normalized eigenvectors of the correlation matrix,
\begin{equation}
\sum_{r,s} K_{ir,js} { \Psi}_{rs} = \mu_k { \Psi}_{rs}^{k} \ \ \ ,
        \hspace{0.5cm} 
     \mu_{1} \ge \mu_{2} \ge \ldots \ge \mu_{m(K+(d-1))} \ge 0 \ \ \ ,
\end{equation}
will now provide an optimal representation of the dynamics of the
spatial modes:
\begin{equation}
{ X}_{ir} (n) = \overset{m(K+(d-1))}{\underset{k=1}{\sum} }
         B_k(n) { \Psi}_{ir}^{k} \ \ \ .
\end{equation}
where
\begin{equation}
B_k(n) = ( {\bf X}(n), {\bf \Psi}^{k}) = 
    \overset{m}{ \underset{i=1}{\sum} } \:
    \overset{K+(d-1)}{\underset{r=1}{\sum} } 
        X_{ir}(n) \Psi_{ir}^{k} \ \ \ .
\end{equation}
We found that $99\%$ of the variance of the input ensemble ($m=2$,
$K=6$ and $d=6$ model) can be represented by the first six temporal
modes ${\bf \Psi}^{k}, k=1,\ldots,6$. We have therefore chosen to
approximate the input vectors presented to the ANN by their six
dimensional POD projection ${\bf B}(n) = (B_1(n), \ldots, B_6(n))$ onto the
space spanned by the first six temporal eigenvectors. Denoting by
${\bf P}_{ {\bf X} \rightarrow {\bf B}}$ this projection operator, the
structure of the model has now become:
\begin{equation}
A_{i}(n+1) = {F}_{i}[ {\bf P}_{ {\bf X} \rightarrow {\bf B}}  
   [ {\bf X}_{1}(n), {\bf X}_{2}(n)] ]
     =   {F}_{i}[B_1(n), \ldots, B_6(n)]
          \hspace{0.3cm}i=1,2 \ \ \ .
\label{final_model}
\end{equation}
The best forecasting performance was obtained for an ANN with two
hidden layers of 10 neurons each.  The input to the network is the
6-dimensional projection ${\bf B}(n)$ and its 2-dimensional output,
$(A_1(n+1), A_2(n+1))$, is the model forecast for the first two mode
amplitudes at the next time step.
Because the large events have long recurrence times, the sequential
selection of the training set will be dominated by input-output pairs
that only express the quasi-linear behavior between two consecutive
large events: the ANN will learn a linear model and will fail to
forecast the intrinsically nonlinear large earthquake events.
Therefore, we have designed the training set to include enough
information describing the dynamics around the scarce large events.
The modal time series were first rescaled to evolve in the interval
$[0, 1]$. We then classified as large events all events in which the
first mode has a jump larger than 0.05 in the rescaled units. Next, we
chose a time window $W=Kd$, and checked for the presence of a large
event located at the center of this window, as it slides along the
modal time series. If we found one, we called this a nonlinear window,
and all the input-output pairs describing the evolution of the system
inside this time window were included in the training set. For each
nonlinear window, we have included training pairs from a linear window
(a segment of the time series that did not have a large
event). With this training set, the ANN adequately learned
to model the linear as well as the nonlinear features of the dynamics.

Due to non-convexity of the error surface in the network parameter
space, we train many ANNs starting from different initializations of
the neuron weights. To avoid overfitting, the training process stops
when the error on the validation set starts to increase - the
validation sample is a subset of the training set which is not
actually used in training. At the end of the training cycle each ANN
provides a model for the system dynamics.  To find the ``best'' model,
we test the forecasting skill of each model realization using time
series segments not included in the training or validation set.
Starting from an initial configuration describing for each input mode
the current amplitude and its past $K+(d-1)$ values, we iterate the
ANN forward in time for a number of $F$ steps. At each time step the
output of the network was used to update and reconstruct the ANN input
for the next time step.  Assuming perfect initial conditions this
procedure provides a trajectory whose forecasting accuracy is
controlled only by the imperfections of our model.  However, the
reconstructed model is always an imperfect one and perhaps a perfect
model describing the large scale motions does not even exists.
Moreover, the current state of the system in the model space is always
obscured by observational uncertainty. To make things worse, our
knowledge of the spatial modes themselves is limited by the finite
size of the ensemble of surface deformations.  For example, the accuracy
of the two-point correlations cannot exceed $1/\sqrt{N}$, where $N$ is
the size of the ensemble: when we have only observations of finite
duration, the ``true'' modes, describing the large scale motions, will
never be exactly known. How can we then evaluate the model
forecasting skill?  A detailed answer is beyond the scope of this
paper and our goal here is only to show that despite all these
difficulties, and many others not mentioned here, our imperfect models
can provide stable and robust forecasts.  Our  remarks here  follow 
Smith \cite{smith00}, to which we also refer for
a clear discussion  of uncertainty in initial
conditions, model errors, ensemble verification, and predictability in
general.

For nonlinear systems, uncertainty in the initial conditions severely
limits the utility of single deterministic forecasts.  Internal
consistency requires that all nonlinear forecast should be ensemble
forecasts \cite{smith00}. In this approach to forecasting, a
collection of initial conditions, each consistent with the
observational uncertainty, are integrated forward in time.  When the
system evolves on an attractor, the selection of the initial conditions
is far from trivial. For a perfect model ( a model that has the right
dynamics and whose phase space is identical with the system's phase
space) the members of the ensemble should be restricted
to live on the system's invariant measure (attractor).  Obviously,
this is not {\it a priori} known. Unrestricted initial conditions,
consistent only with the observational uncertainty, will extend into
the full phase space and generate over-dispersive ensembles. When this
is the case, the predicted probabilities will not match the relative
frequencies as demonstrated by Gilmour \cite{gilmour}. Generally, all
models, including ours, are imperfect and, therefore, a perfect
ensemble and an accountable forecast method do not
exists. Nevertheless, when the initial conditions are uncertain,
ensemble forecasts are still required.
A reasonable constraint
would require the members of the ensemble to live on the projection of
system's invariant measure (attractor) into the model phase space, although
this choice is not necessarily optimal or even unique \cite{smith00}.

Assuming no {\it a priori} knowledge of the true attractor projected
into the model phase space, we present in Fig. \ref{forecast} an
unconstrained ensemble of trajectories evolving under the dynamics of
the ANN model. Each trajectory starts from a different initial state
that contains small Gaussian perturbations from the exact initial
conditions and is integrated forward in time for $F=250$ steps. We
compare the ensemble evolution (blue lines) with the true evolution of
the most dominant surface mode (red, thick line). 
The resulting
ensembles of forecasts can be interpreted as a probabilistic
prediction.  The goal is to predict the time and the size of the jumps
in the evolution of the first mode amplitude (which, as we have
already discussed, corresponds to the time and size (potency) of large
seismic events), and to estimate their forecast accuracy.  Due to the
time delay involved, the best time resolution of each trajectory
cannot be in this case less than $\Delta t=0.1$ yr.  As the fault
evolves to the next time step, $0.1$ years later, we update the
present state of the system and generate a new $F$ step ensemble
forecast starting from this new state. This procedure is intended to
incorporate the information about the system and its current state as
is continuously generated by new observations.

Is clear from Fig. \ref{forecast} that even though we lack a perfect
model or an optimal ensemble, all members of the ensemble forecast the
incoming sequence of large events.  The forecast uncertainty of the
ensemble grows very slowly showing long time reliability.
The members of the ensemble spread out at a rate that depends on the
local nonlinear structure of the model. This rate gives a local
estimate of the stability of forecasts made in this region of the
model's state space.  It also controls the time scale (predictability
time) on which the ensemble members scatter along significantly
different trajectories.  We observe regions of large predictability time
that coexist with regions of relatively short predictability time
\cite{bcfv02}.
Significantly, the ANN has shown consistently the ability to improve
its forecasting as the system approaches a large event. The
predictability time is significantly larger than the microscopic
expansion rate which is controlled by the maximal Lyapunov
exponent. As hinted earlier, the predictability is scale dependent and
the large scale motions are more predictable than the small scale
motions.

The predictability also depends on the location of the system on its
true underlying attractor. But unlike the growth of initial
uncertainty in model phase space, the local nonlinear structure of the
system's attractor controls how rapidly the small scale motions, not
included in our model, progress to reach the large length scales. This
expansion rate defines how rapidly the {\it truth}, red line in
Fig. \ref{forecast}, diverges from the best guess model
trajectory. This is the trajectory starting from the true system state
projected onto the model phase space. Of course, for the same model
state there are infinitely many system states distinguishable only in
their small length scale structure. Clearly, the accurate shadowing of
the {\it truth} by the ensemble trajectories in Fig. \ref{forecast},
shows that growth of the small scale uncertainties does not severely
limits the model forecasting skill. Their small growth rate makes
possible the deterministic modeling of the large scale motions.  It
sets an upper bound for the model predictability time, which can be
approached by improving the parametrization of the small scale dynamics.

\section{8. Conclusions}

We describe a conceptual framework for modeling and forecasting the
evolution of a large strike-slip earthquake fault.  The approach
relies on the detection of spatio-temporal strain patterns embedded in
the {\it observable} surface displacements: no detailed knowledge of
the fault geometry, dynamics, or rheology is required.  Rather than
 directly modeling the fault dynamics, we propose instead to
model the dynamics of observable surface deformations, which are
nonlinearly related to the original dynamics of the fault system.

The essence and novelty of the method lies in the discovery that the large
length and time scales dynamics have a strong
{\it low-dimensional, deterministic} component and are therefore
amenable to representation by a deterministic  model. We have
also found that the large scale motions provide reliable forecasting
information about the large seismic events. These two fundamental
results set the stage for standard data processing and model
reconstruction techniques.  First, we identify the large scale motions
with generalized directions (spatial modes) along which the dynamics
has its {\it largest} fluctuations. Finding these directions is the
natural task of the proper orthogonal decomposition applied to the
ensemble of surface deformations generated during the evolution of the
system.  The most dominant spatial modes define the model phase space
and provide an optimal embedding for the large scale dynamics of the
system.  Second, the model reconstruction consists in finding a
nonlinear set of ODEs whose trajectory in model phase space
approximates the system trajectory projected into the model phase
space. This is a {\it learning} task that can be successfully
accomplished by an artificial neural network.

The method relies on the existence of some separation between small
and large length and time scales, and the physics responsible for this
separation stems from the dynamic weakening used to model static/kinetic
friction. We argue that in the presence of significant dynamic
weakening the large scale dynamics defines the low-dimensional
backbone of the system's attractor. The small scale dynamics is
practically indistinguishable from stochastic dynamics and evolves in
a small neighborhood of the attractor backbone.  Based on this
geometric picture, we argue that the statistical properties of the
small scale motions are determined by the larger-scale motions upon
which they are superposed. In turn, the large scale motions depend on
the statistical properties of the small scales. As pointed out by
Lorenz \cite{lorenz69}, there remains uncertainties in the small scale
statistics, and hence in their influence upon the larger scale. The
predictability time of the large scale motions is controlled by the
rapidity with which the small scales uncertainties progress to reach
the very large scales.  This sets an upper bound to the predictability
time of any large scale model. Due to the robustness shown by the
model ensemble forecast, we conclude that the intrinsic predictability
time of the large scale dynamics is very long, perhaps of the order of
10-20 years. This is ultimately the reason behind the deterministic
behavior of the large scale motions.

There are many difficult problems that are currently under
investigation. For example, we presently study how robust the large
scale behavior is to changes in the model parameters. Preliminary
results confirm the controlling role of the dynamic weakening and show
that the deterministic character of the large scale motions is robust
to large changes in this parameter. We are also studying the effects
of correlated heterogeneities, smooth brittle to ductile transitions,
and continuous transition from static to kinetic friction. We do
expect that all these effects are averaging out the small scale
dynamics and strengthen the deterministic structure of the attractor.
Probably, the most difficult problems to address are the construction
of good models from short data observations and the problem of
spatio-temporal chaos. In the last case, dimensions and Lyapunov
exponents become intensive quantities and we want to understand how
the method scales with the size of the fault system. The current paper
is concerned primarily with developing a methodology and the obtained
results are preliminary. Continuing studies along the directions of
this work may have a significant impact on the earthquake
predictability problem.

\section*{Acknowledgments}

\noindent
{\small We express our gratitude to Yannis Kevrekidis for useful discussions 
 and insightful suggestions. 
This research was performed under the auspices of the U. S.
  Department of Energy at LANL (LA-UR-02-6309) under contract
  W-7405-ENG-36 and LDRD-DR-2001501 (MA) and the National Earthquake
  Hazard Reduction Program of the USGS under grant 02HQGR0047 (YBZ).
  }

\pagebreak

\section*{Captions}
\begin{figure}
  \centering\epsfig{file=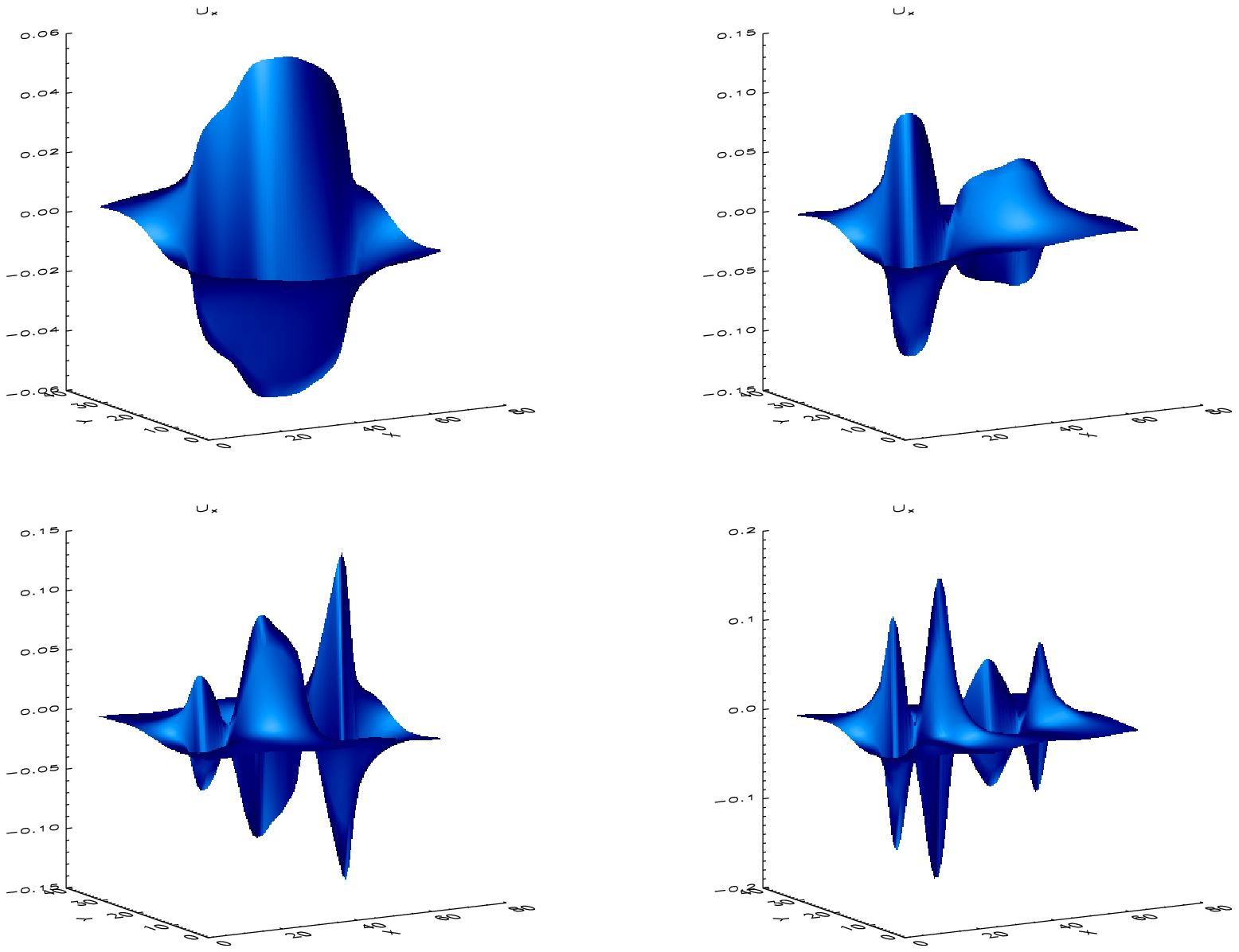,height=100mm,width=150mm}
  \caption{} \label{ux_evecs}
\end{figure}

\begin{figure}
  \centering\epsfig{file=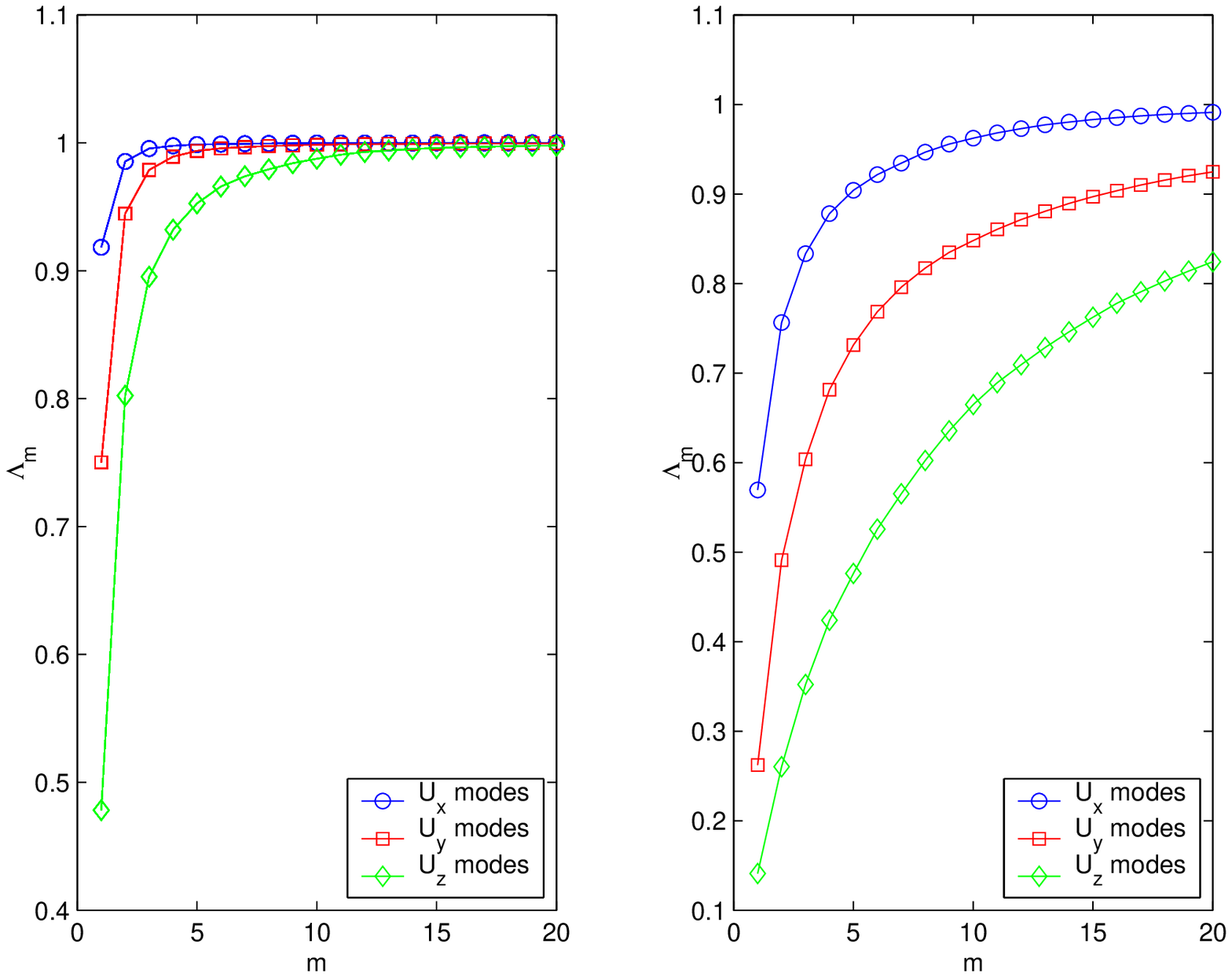,height=60mm,width=150mm}
  \caption{} \label{cum_evals}
\end{figure}

\begin{figure}[y]
  \centering\epsfig{file=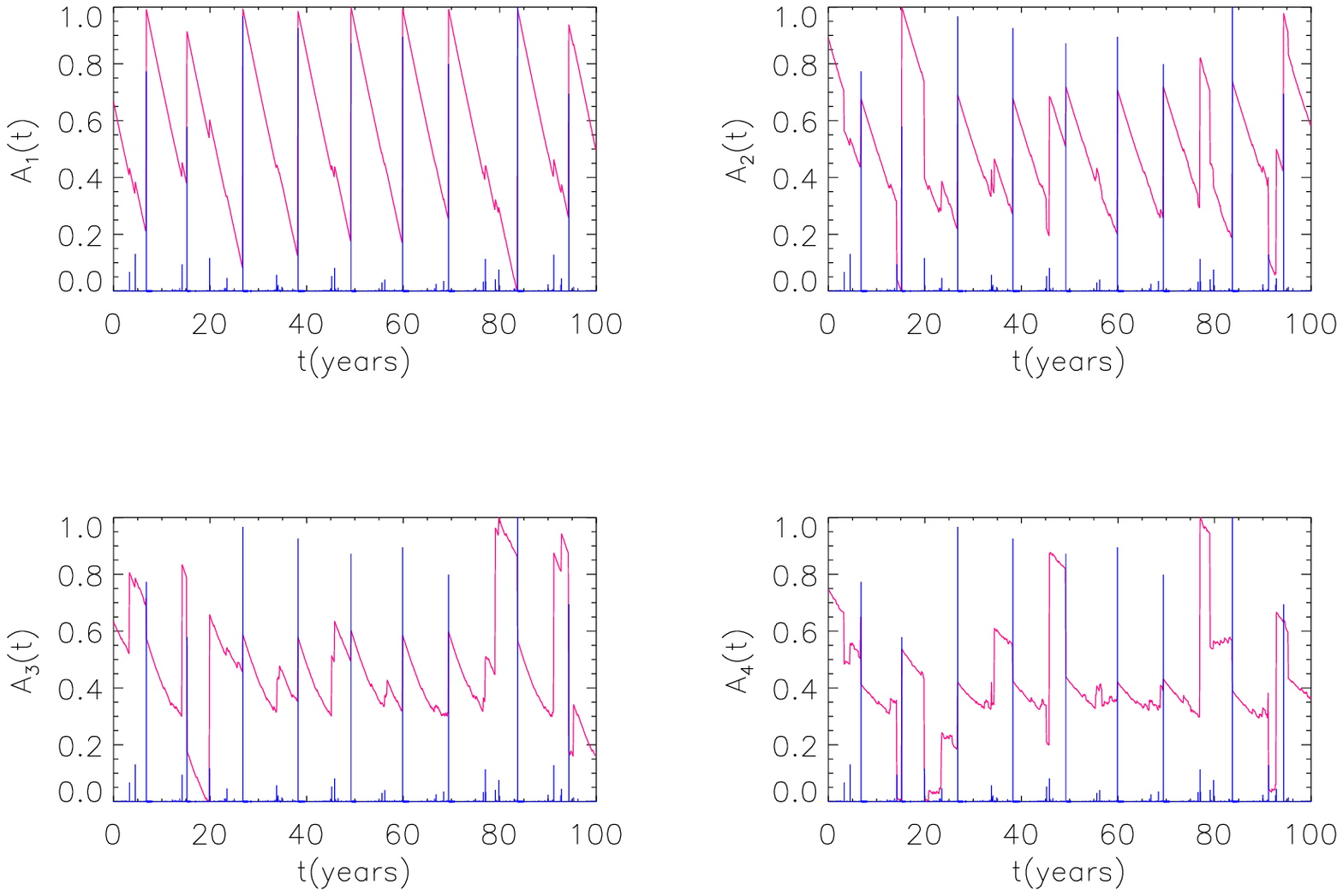,height=100mm,width=100mm}
  \caption{} \label{ux_amp_evol}
\end{figure}

\begin{figure}[b]
  \centering\epsfig{file=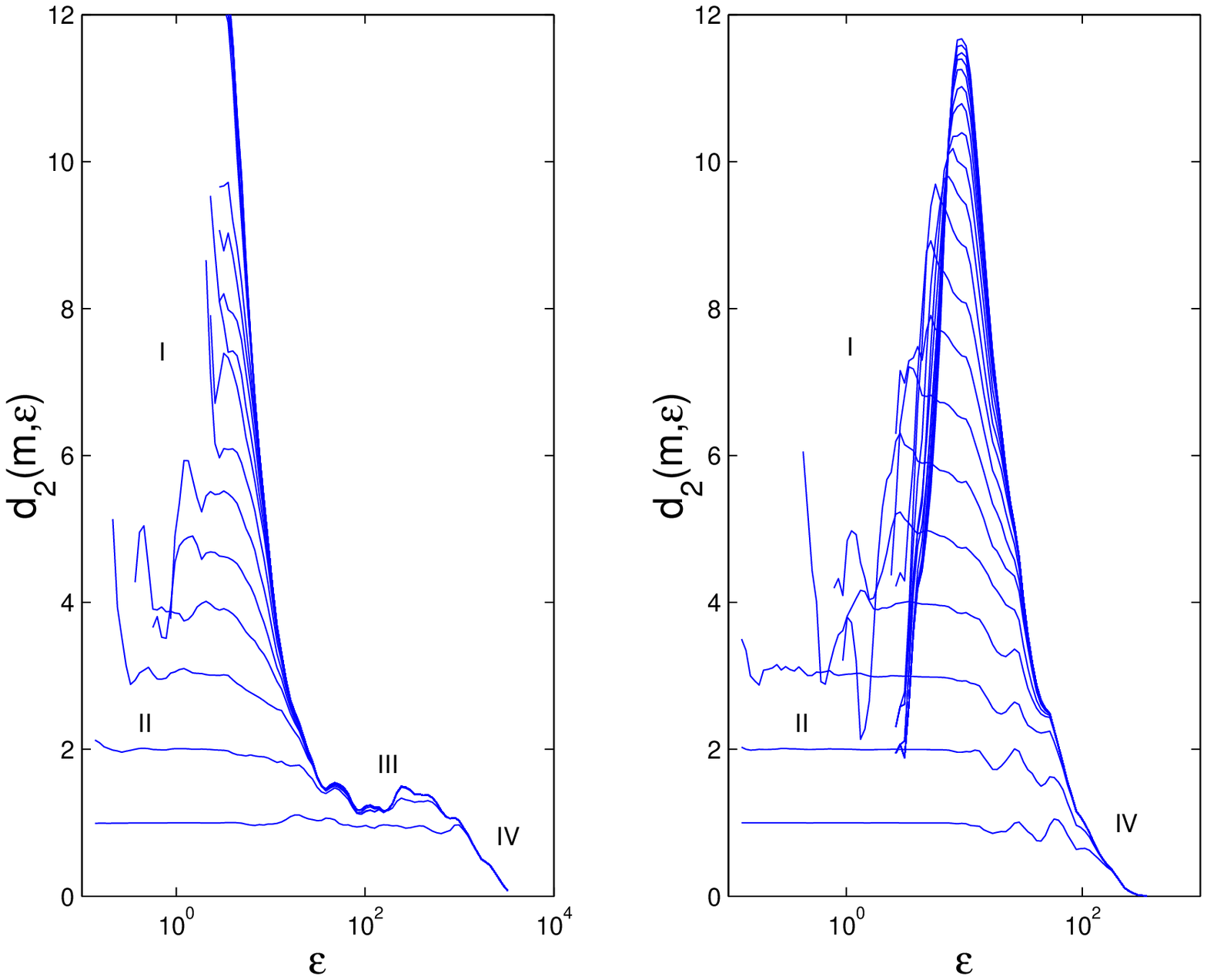,height=60mm,width=150mm}
  \caption{} \label{d2}
\end{figure}

\begin{figure}[t]
  \centering\epsfig{file=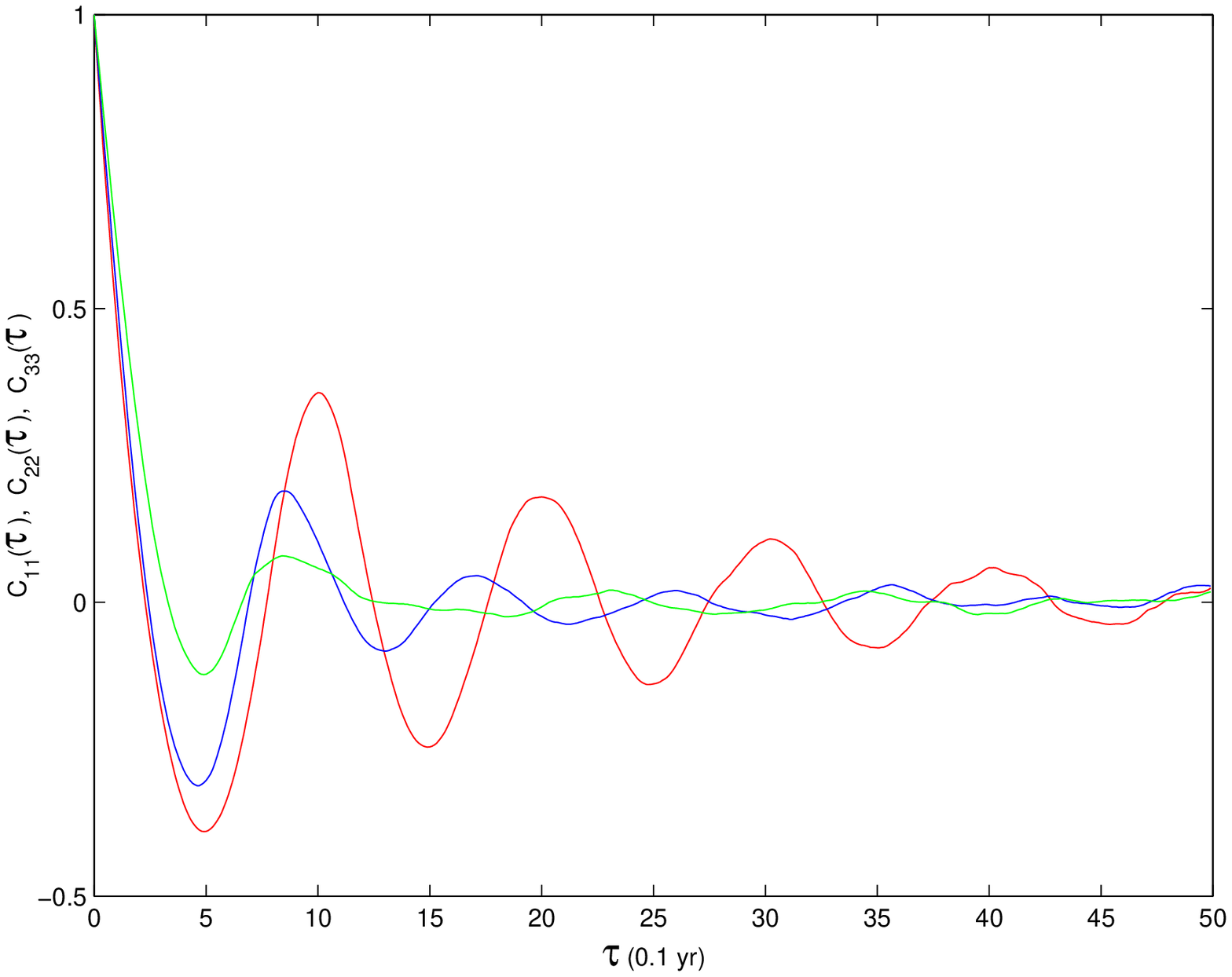,height=60mm,width=80mm}
  \caption{} \label{time_corr}
\end{figure}

\begin{figure}[b]
  \centering\epsfig{file=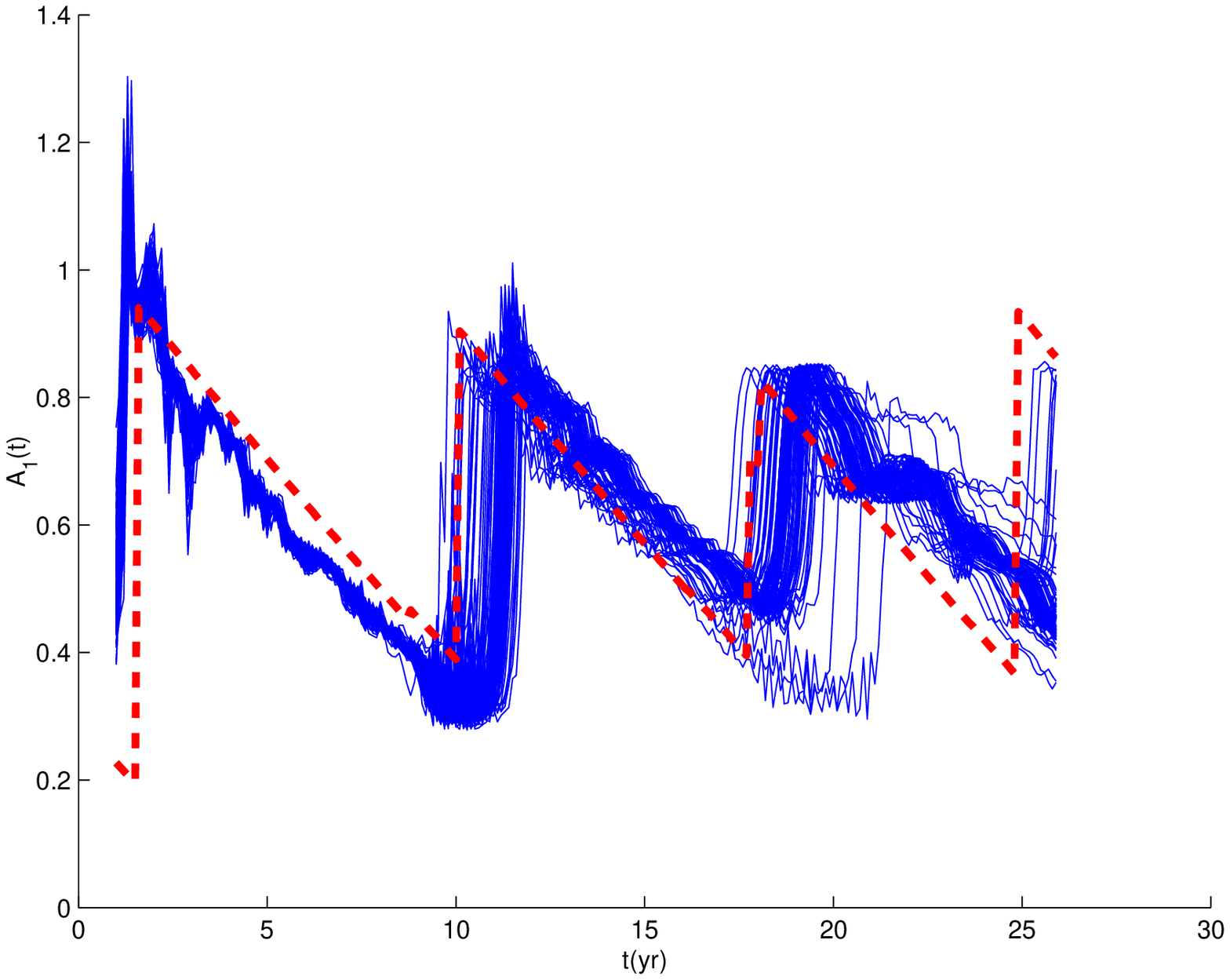,height=60mm,width=80mm}
  \caption{} \label{forecast}
\end{figure}

\begin{itemize}
\item[Fig. 1] Schematic evolution of the stress and slip for
an individual fault cell. The parameters that describe the
static/kinetic friction law are: static strength, $\tau_s$, dynamic
strength, $\tau_d$, and arrest stress, $\tau_a$. The time dependent
failure strength of the cell, $\tau_f$, equals the static strength if
the cell has not slipped yet and is adjusted to its dynamic value when
the cell fails (from \cite{ybz93}).

\item[Fig. 2]  A $30$ years evolution of the potency released
by two fault systems. Results for system A, with small dynamic
overshoot $D=1.5$ are on the left and for system B, with large dynamic
overshoot $D=\infty$ (and no dynamic weakening) are on the
right. System A exhibits quasi-regular seismic cycles associated
with the large events. There is no detectable
structure in the evolution of seismicity in system B.

\item[Fig. 3] Time evolution of the average
slip deficit (left) and its variance (right) for the $D=1.5$ (blue
line) and $D= \infty$ (red line) fault systems.

\item[Fig. 4]  Log-log frequency-size histogram plots for the
$D=1.5$ system (blue squares) and the $D=\infty$ system (red
circles). The earthquake size is measured by its potency and the
histograms use equal bins of size 100 in potency units ($0.29
\mbox{mm}\times\mbox{Km}^2$). 

\item[Fig. 5] The fraction of false nearest neighbors (FNN)
as a function of the embedding dimension $m = 1, \ldots,10$. Compared
to model B ($D= \infty$), there is a dramatic decrease in the number of FNN
for model A ($D=1.5$), as the dimension of the embedding phase space
is increased. This suggests that a low-dimensional deterministic dynamics
is a good approximation for the large scale dynamics of the fault.

\item[Fig. 6] Estimates of the maximal Lyapunov exponent from the slip
deficit variance time series data, $\sigma(t)$, for the $D=1.5$
model. The logarithm of the stretching factor is computed for three
different neighborhood sizes, $\epsilon$, and $m=6,7,8,9,10$ embedding
dimensions. The robust linear behavior of $S(\Delta n)$ reflects the
underlying determinism of the data and its slope is an estimate of the
maximal Lyapunov exponent, $\lambda_{max} = 0.8\mbox{\,\,years}^{-1}$ (the
straight line has slope $0.08$ and the unit of time is $0.1$ years).

\item[Fig. 7] The first four POD spatial modes of the
$u_{x}$ surface deformation field for fault system A with dynamic
overshoot $D=1.5$. These modes define, in decreasing order of their
eigenvalues, generalized deformation directions which support most of
the variance produced by the dynamics of surface fluctuations. The
coherent, collective nature of the modes is reflected in their overall
shape.

\item[Fig. 8] Cumulative normalized eigenvalue spectrum for each of
the surface deformation modes for system A on the left and B on the
right - only the first 20 POD modes are shown. In order to explain
$f=95\%$ of the variance in the $u_{x},u_{y},u_{z}$ data sets for
system A we need only 2, 3, and 5 modes respectively. In contrast, for
system B we need 9, 27, and 42 modes.  Generally, the surface
deformations along the strike direction, $u_{x}$, have the most
efficient POD description.

\item[Fig. 9] Modal time series (red lines) describing the
evolution of the first four $u_{x}$ spatial modes shown in
Fig.~\ref{ux_evecs}. This time dependence is determined by projecting
the $u_{x}$ surface deformation onto each POD mode
every $0.1$ years. The blue lines describe the time evolution of the
 cumulative potency released over
each $0.1$ year interval. Note the correlations between the time
intervals of high potency released and the discontinuities present in the
temporal modes.

\item[Fig. 10] Local slopes $d_{2}(m,\epsilon)$ of the correlation sum
for the multivariate time series describing the evolution of systems A
(on the left) and B (on the right) for embedding dimensions
$m=1,\dots,20$. For system A ($D=1.5$), we detect the emergence of a
scaling region around $\epsilon \approx 100 $, suggesting a
correlation dimension $d_{2} \simeq 1.3$ and self-similar geometry of
the attractor at the large length scales. This behavior breaks down
when the attractor is probed at smaller length scales where the
dynamics is high dimensional. Within the same range of embedding
dimensions, system B ($D=\infty$) shows no signature of a large scale
dimension collapse.

\item[Fig. 11] The temporal autocorrelations of the first three modes
of the $u_x$ deformation field in $D=1.5$ model (mode 1 red line,
mode 2 blue line, and mode 3 green line). The short-term oscillatory
behavior and the slow long-term amplitude decay reveal the presence of
two correlation time scales. The oscillations become faster and the
long-term coherence decreases rapidly for the higher order modes.

\item[Fig. 12] Model ensemble forecast for the evolution of
the first $u_{x}$ mode (coordinate $A_1$ In model phase space)
as predicted by the iterated ANN (blue lines).  Each trajectory of the
ensemble starts from a different initial condition and is integrated
forward in time for 250 steps.  Each ensemble trajectory evolves into
the model phase space, while the red dashed line represents the true
trajectory of the system projected onto the first coordinate of the
model phase space (mode one). The ensemble is consistent with an
arbitrary Gaussian, initial observational uncertainty, but its members
do not live on the projection of the true system's attractor 
into the model phase space. Nevertheless, all members of the ensemble
consistently forecast the incoming sequence of large events, proving
the long term forecasting reliability.  We also notice a systematic
bias in estimating the time to the next large event. 
\end{itemize}

\pagebreak



\pagebreak


\pagebreak


\pagebreak


\end{document}